\documentclass[preprint]{aastex63}

\renewcommand{\thefootnote}{\fnsymbol{footnote}}
\usepackage{mathtools}
\usepackage{cleveref}
\usepackage{soul}

\graphicspath{ {./Figures/} }

\received{December 6, 2021}
\revised{June 26, 2022}
\accepted{September 1, 2022}
\shorttitle{VLA + Juno}

\begin{document}
\title{Ammonia Abundance Derived from Juno MWR and VLA Observations of Jupiter}

\correspondingauthor{Chris Moeckel}
\email{chris.moeckel@berkeley.edu}

\author[0000-0002-6293-1797]{Chris Moeckel}
\affiliation{University of California, Berkeley\\
Department of Earth and Planetary Science \\ 
307 McCone Hall, \\
Berkeley, CA 94720, USA}

\author[0000-0002-4278-3168]{Imke de Pater}
\affiliation{University of California, Berkeley \\ 
Department of Astronomy}

\author[0000-0003-3197-2294]{David DeBoer}
\affiliation{University of California, Berkeley \\ 
Department of Astronomy}




\begin{abstract}
The vertical distribution of trace gases in planetary atmospheres can be obtained with observations of the atmosphere's thermal emission. Inverting radio observations to recover the atmospheric structure, however, is non-trivial, and the solutions are degenerate. We propose a modeling framework to prescribe a vertical distribution of trace gases that combines a thermo-chemical equilibrium model {based on a vertical temperature structure and compare these results to models where ammonia can vary between pre-defined pressure nodes}. To this means we retrieve nadir brightness temperatures and limb-darkening parameters, together with their uncertainties, from the Juno Microwave Radiometer (MWR). We then apply this framework to MWR observations during Juno's first year of operation (Perijove passes 1 - 12) and to longitudinally-averaged latitude scans taken with the upgraded Very Large Array (VLA) \citep{dePater2016,dePater2019}. We use the model to constrain the distribution of ammonia between -60$^{\circ}$ and 60$^{\circ}$ latitude and down to 100 bar. We constrain the ammonia abundance to be $340.5^{+34.8}_{-21.2}$ ppm  ($2.30^{+0.24}_{-0.14} \times$ solar abundance), and find a depletion of ammonia down to a depth of $\sim$ 20 bar, which supports the existence of processes that deplete the atmosphere below the ammonia and water cloud layers. At the equator we find an increase of ammonia with altitude, while the zones and belts in the mid-latitudes can be traced down to levels where the atmosphere is well-mixed. The latitudinal variation in the ammonia abundance appears to be opposite to that shown at higher altitudes, which supports the existence of a stacked-cell circulation model.  
\end{abstract}



\section{Introduction} \label{sec:intro}

Some of Jupiter's biggest mysteries remain unsolved despite decades of observations. Questions about its formation environment, the source of energy for its atmosphere, and the distribution of trace gases within said atmosphere remain as open-ended as ever. The vastness of the planet and its atmosphere, coupled with the short time scales on which the atmosphere evolves, implies that a global understanding of the planet requires coordinated observations across many different frequencies. In the early days of observational astronomy, optical images were used to infer the dynamics of the atmosphere by tracking the motion of cloud tops \citep{peek_1959,Ingersoll1981}. Various techniques were used to correlate images and find the dominant zonal component of the global circulation. The meridional component of the atmosphere is much smaller in magnitude, making it more difficult to observe. Only recent advances in observational capabilities and computing techniques have allowed for direct measurement of the north-south component of the wind by tracking small scale cloud features \citep{Asay-davis2011}. 

A key feature in expanding this limited two-dimensional view of the atmosphere is observations of the planet's thermal emission at infrared \citep{Gierasch1986} and radio wavelengths. At radio wavelengths the impact of clouds is sufficiently small such that we can still draw conclusions about the atmosphere below, allowing radio observers to speculate as early as the 1960's about the presence of microwave absorbers in the Jovian atmosphere below the visible cloud tops \citep{Thornton1963,Welch1966,Wrixton1971,Berge1976}. Given the right combination of frequencies, we can use radio observations to construct a full three-dimensional view of the planet and reveal global circulation patterns.

Studying this pattern on the planet requires observations at resolutions high enough to distinguish between the various climatological zones, i.e., the visibly white zones and brown belts on Jupiter. Ground-based observatories use the ability to detect phase at radio frequencies and rely on complex interferometers, with baselines on order of 10s of kilometers, such as the Very Large Array (VLA) and the Atacama Large (sub)Millimeter Array (ALMA), both providing excellent spatial resolution from Earth. The NASA Juno mission opted for an alternative solution, building an orbiter that approaches Jupiter as close as 3000 kilometer. Intriguingly, despite the fact that Juno is five orders of magnitude closer to Jupiter than the VLA, the final resolution at 10 GHz or higher is comparable on the planet when the VLA is in its extended configuration as shown in \Cref{fig:PJ3_map}. Though both approaches have orthogonal advantages and disadvantages for atmospheric observations, the combination provides a very powerful tool for studying the atmosphere of Jupiter. The ground-based observations have excellent spectral and spatial resolution (both in latitude and longitude), but are limited by the presence of Jovian synchrotron radiation along the line of sight at frequencies below ~5 GHz, which hinders the VLA from retrieving information below $\sim$ 10 bar. The proximity of Juno to the planet allows for observations that drastically reduce the impact of synchrotron radiation; however, data are taken along a narrow north-south track so there is {limited global context} for these observations. 

The model we develop in this work is applicable to both sets of measurements, and provides a reference frame for fitting and interpreting radio observations. In Section \ref{sec:data}, we present our data reduction pipeline developed for the Juno data
, followed by our description of the perturbed thermo-chemical equilibrium model for the atmospheric structure in Section \ref{sec:methods}. Based on a parameterized distribution of trace gases, we can compute the brightness temperature using our radiative transfer code, as described in Section \ref{sec:RT}. In Section \ref{sec:results}, we present our results that minimize the difference between the observations and the model atmosphere for the mean atmosphere based on the first year of Juno observation, and compare our results to the only other published result for perijove (PJ) 1. Lastly, we touch upon the most important interpretations in Section \ref{sec:discussion}. We finish the paper with our overall conclusions and recommendations for the future work.

\section{Data} \label{sec:data} 
\subsection{Sources of Data} \label{ssec:dj}
Since Juno's insertion into Jupiter's orbit on July 4th, 2016, the spacecraft has remained in a 53 day, highly eccentric, polar orbit, in which the spacecraft spends most of its time outside of the high radiation environment around the planet. Among the suite of instruments, there are several with the primary goal of studying the planet's atmosphere, including a microwave radiometer \citep[MWR; ][]{Janssen2017}. This instrument couples six radio receivers to a set of antennas (an array of patch antennas and corrugated horns) to receive radio waves with minimal side lobe contributions at six distinct frequencies, from here on referred to as Channel 1 through 6 (see \Cref{tab:mwrfreq} for details).

\begin{table}[h]
\centering
\caption{Radiometer and antenna characteristics of Juno's MWR.}
\begin{tabular}{c|c|c|c|c|c|c}
Channel             & 1    & 2     & 3     & 4     & 5      & 6    \\ 
Frequency [GHz] & 0.6  & 1.278 & 2.597 & 5.215 & 10.004 & 22   \\ \hline
Bandwidth [MHz] & 21   & 43.75 & 84.5  & 169   & 325    & 770  \\ \hline
Beamwidth [deg] & 20.6 & 20.6  & 12.1  & 12.1  & 12.0   & 10.8
\end{tabular}
\label{tab:mwrfreq}
\end{table}

Jupiter's emission at radio wavelengths consists of two components: a thermal component (i.e., the blackbody radiation of the planet's atmosphere) and non-thermal emission originating from synchrotron radiation by electrons in the radiation belts, sferic emission by lightning on the planet and auroral emission \citep[e.g., ][]{Berge1976,dePater1990,Zarka1998,Brown2018}. As seen from Earth, part of the synchrotron radiation is spatially co-located with the atmospheres' thermal emission \citep[see, e.g., ][]{dePater1986}, which makes observations increasingly difficult below 7 GHz (the frequency at which synchrotron radiation becomes dominant \citep{Berge1976}). Juno circumvents this superposition of emitters by dipping between the planet and the radiation belts during PJs, which brings the spacecraft as close as $\sim$3000 km above the cloud tops of the planet. \\ 

The spacecraft completes one rotation every 30 seconds, in which time its beams sweep across the sky (sky coordinates $\theta_{sky},\phi_{sky}$) and obtain 300 beam-convolved measurements (i.e. the brightness temperature $T_b$) of the sky's brightness ($T_{sky}$) at positions expressed in spherical coordinates in the spacecraft frame  ($\theta_{sc},\phi_{sc}$) \citep{Janssen2017}: 

\begin{equation} 
T_b = \int_{0}^{\pi}\int_{0}^{2\pi} {T_{sky}(\theta_{sky},\phi_{sky}) G(\theta_{sky}-\theta_{sc},\phi_{sky}-\phi_{sc})} sin(\theta_{sc}) d\theta_{sc} d\phi_{sc}
\label{eqn:TA}
\end{equation} 

The radiation pattern (or `gain') of the radio antenna $G(\theta,\phi)$ is designed so that 99\% of the emission falls within 2.5 times the half power beam width (HPBW) of the Gaussian main beam, limiting the amount of radiation received through the side lobes. The intersection of the main beam with the planet determines the resolution of the measurements, which is highest at nadir viewing, and decreases with increasing slant angle, distance from the planet, and frequency. For ideal operation of the MWR, the antennas scan the planet in a north-south direction to obtain multiple measurements per latitude at varying emission angle (the angle between the antenna's line of sight and the surface normal). While measurements at nadir probe deepest into the planet, observations at increasing emission angles scan the planet higher up in the atmosphere due to the increased path lengths. Juno's other science priorities, such as gravity science, require communication antennas to be aligned with Earth so that the satellite does not align with the spin axis of the planet, which deteriorates the emission angle coverage { and reliability of the atmospheric signal (such as PJ 10 and 11), which we therefore exclude from this analysis}. The most relevant information for the MWR observation campaign are listed in \Cref{tab:JunoCampagain}. 

The radiation pattern for the six antennas, the spacecraft geometry during PJs, and the corresponding antenna temperature measurements are documented in the Planetary Data System\footnote{Dataversion 3, \url{https://pds.nasa.gov/}} and are utilized in the following analysis.  \\


\begin{table}[]
\centering
\caption{MWR Orbits used in our work}
\begin{tabular}{c|c|c|c|c}
PJ number & Date       & Min. dist (km) & PJ {[}lat,lon{]}               & Lon range  \\ \hline
1         & 2016-08-27 & 4138           & 4.3, 95.8                            & -115, -75  \\ \hline
3         & 2016-12-16 & 4101           & 6.5, - 5.5                           & -18, 5     \\ \hline
4         & 2017-02-02 & 4234           & 7.5, 85.1                            & 64, 107    \\ \hline
5         & 2017-03-27 & 3314           & 8.6, 175  1                          & 64, 107    \\ \hline
6         & 2017-05-19 & 3383           & 9.7, -140                            & 165, -177  \\ \hline
7         & 2017-07-11 & 3359           & 10.8, -49.0                          & -156, 114  \\ \hline
8         & 2017-09-01 & 3332           & 11.8, 41.9                           & -73, -21   \\ \hline
9         & 2017-10-24 & 3839           & 13., 132.                             & 30, 73    \\ \hline
12        & 2018-05-24 & 3201           & 15.9, 115                            & -144, -107 \\ 
\end{tabular}
\newline
\newline
Overview of MRW's orbits which yield data useful for atmospheric science. The columns correspond to the PJ number, date of the flyby, minimum distance from the planet, sub satellite location during closest encounter, and the longitude range covered during the flyby.
\label{tab:JunoCampagain}
\end{table}

In addition to the Juno data, we also incorporate observations between 3 and 25 GHz from the Very Large Array (VLA) into this analysis. These were originally published in \citet{dePater2016} and \citet{dePater2019}, and we refer the reader to these papers for details on the data reduction and calibration.   \\

\subsection{Data reduction of the Juno observations} \label{ssec:dr}
The two quantities needed from the Juno data for our atmospheric retrieval (see Section \ref{sec:results}) are the nadir brightness temperature at the observing frequency, $T_{b0}$, and the limb-darkening coefficient $p$, which numerically represents how the brightness temperature decreases at increasing emission angle $\theta$ ($\mu \equiv \cos \theta$) as we probe higher up in the atmosphere compared to nadir observations:

\begin{equation}
    T_b(\mu)  = T_{b0} {\mu}^{p}
    \label{eq:TBp}
\end{equation}

\noindent with $T_b(\mu)$ the brightness temperature of Jupiter at a given frequency convolved with the gain pattern of the antennas (\Cref{eqn:TA}).

\begin{figure*}[h]
\centering
\includegraphics[width=\textwidth,trim={6cm 0cm 6cm 0},clip]{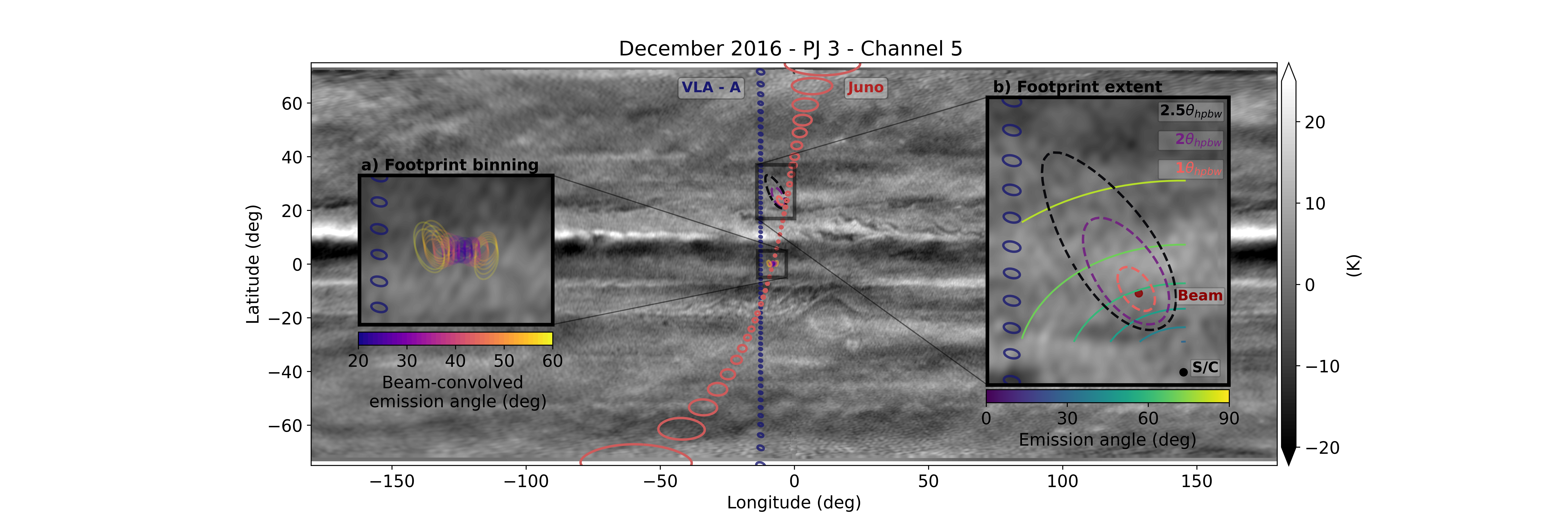}
\caption{PJ3 flyby during December 2016. The map shows X-band (8-10 GHz) VLA observations taken concurrently with the flyby to show the longitudinal structure of the planet \citep[][These VLA data will be fully published in a subsequent paper]{Moeckel2019}. The red footprints correspond to MWR's Channel 5 observation, where we have plotted the highest resolution (i.e. smallest emission angle), while the blue footprints correspond to the VLA resolution (Note: In December 2016 the VLA was in the most extended A-configuration). Inset a) is a zoom-in for the indicated region and shows all the 60 individual observations colorized by their beam-convolved emission angle that are centered around the equator. These 60 observations are the basis for fitting the nadir brightness temperature and the limb-darkening at 0$^{\circ}$. Inset b) is a zoom-in to illustrate how the emission angle varies across the beam, where the red central dot corresponds to the boresight center and the black dot indicate the position of the spacecraft. The dashed ellipses correspond to the beams' extent projected onto the planet with the contours showing how the emission angle varies across the beam. The final beam-convolved emission angle is 59.2 deg when integrating over the 2.5 $\theta_{HPBW}$ contour that corresponds to 99$\%$ of the emission.}
\label{fig:PJ3_map}
\end{figure*}

For a full treatment of the problem, the observations can be deconvolved based on an exact knowledge of the antenna pattern and using several overlapping observations \citep{Oyafuso2020}. We choose an alternative approach, and instead correct the observations for the viewing geometry. Since we are interested in obtaining the variation in brightness temperature with emission angle, we must account for the fact that the beam captures a wide range of emission angles. Based on the shape of the planet and the relative position of the antenna beam, we derive a map of emission angle for each observation (see inset in \Cref{fig:PJ3_map}, and Appendix A). Knowing the detailed shape of the antenna pattern, we can convolve the two quantities to obtain a beam-averaged temperature ${T_b(\bar{\mu})}$. Assuming that {the variations in the observable quantities (i.e. nadir brightness temperature and limb-darkening) } across the beam are small compared to variations in brightness due to changes in emission angle, we obtain the final brightness temperature of the observation as the mean brightness temperature across the beam ${T_b(\bar{\mu})}$ at the mean emission angle across the beam $\bar{\mu}$:

\begin{equation} 
    T_b(\bar{\mu}) = \bar{T}_{b0} \bar{\mu}^{\bar{p}}
    \label{eq:Tld}
\end{equation}

\begin{equation}
    \bar{\mu}  = \frac{\iint_{0}^{\theta_{sc}}G(\theta_{sc},\phi_{sc})\mu(\theta_{sc},\phi_{sc})}{\iint_{0}^{\theta_{sc}}G(\theta_{sc},\phi_{sc})}
    \label{eq:mubar}
\end{equation}

We note here that other parameterizations for the limb-darkening can be used \citep[such as the two term expansion by ][]{Li2017}); however, the exact form does not influence the atmospheric retrieval. All we care about for the retrieval is by how much the brightness temperature has dropped at an emission angle of 45$^{\circ}$ for our model. 

We set the limits of the beam convolution equal to 2.5 times the HPBW of the beam, corresponding to the angle within 99\% of the total emission is received \citep{Janssen2017}. In case a beam element falls off the planet, we chose to discard that measurement due to drastic brightness contrast between the planet and either the background sky or synchrotron radiation.\\ 

{To derive the latitudinal structure of the observables, we first collect all measurements were the boresight falls within a 1 deg latitude bin. The resolution of the actual observations is set by the size of the footprints and as such varies between each measurement, often larger than the 1-degree bin size. This bin size, therefore, compromises between resolving the latitudinal structure and the size of the footprints. Inset a) in \Cref{fig:PJ3_map} shows an example of this, where we plot all 60 observations where the beam-center falls within the selected region, and the beams at large emission angles indeed span a few degrees of latitude. 
To that purpose we fit all data within the bin using a weighted linear least squares (\Cref{eq:wlls}) based on parametric relationship between $T_{b0}$ and $p$ (see \Cref{eq:Tld}). At the lowest frequencies (Channel 1 - Channel 3) we suspect contamination by synchrotron radiation to be largest, so that use $\bar{\mu}$ as a weight (${W}$). The choice of $\bar{\mu}$ only slightly downweighs the observations ($\mu(\theta = 45) \approx=0.7$), and as such presents a compromise. If we used weights that penalize large emission angles even more, we would lose important information on the limb-darkening, while a smaller weight caused problems at mid latitudes with synchrotron radiation leaking in.}

\begin{equation} 
    \frac{\delta}{\delta X} \sum\limits_{i=1}^n W({Y}-{A}{X})^2 = 0 
    \label{eq:wlls}
\end{equation}

\begin{align*}
W =  \begin{bmatrix} 
    \bar{\mu_1}&  &\hdots &   & 0\\ 
    \vdots & \ddots & & & \vdots  \\ 
    0  &  \hdots&  \bar{\mu_i}& \hdots&  0  \\ 
    \vdots &  & &\ddots &  \vdots\\ 
    0 & & \hdots &  & \bar{\mu_n}\\ 
  \end{bmatrix} 
&& 
Y =  \begin{bmatrix}
    log T_b(\bar{\mu_1}) \\  
    \vdots \\ 
    log T_b(\bar{\mu_i})  \\ 
    \vdots  \\ 
    log T_b(]bar{\mu_n}) 
  \end{bmatrix}
&& 
A =  \begin{bmatrix}
    1 & log \bar{\mu_1}\\ 
    \vdots & \vdots \\ 
    1 & log \bar{\mu_i} \\ 
    \vdots & \vdots \\ 
    1 & log \bar{\mu_n} 
  \end{bmatrix}
&& 
X =  \begin{bmatrix}
  log T_{b0}  \\ 
  p 
  \end{bmatrix}
\end{align*}

Based on best fit parameters we compute the error $\epsilon$ by subtracting the data from the parametric fit, and compute the reduced chi-squared value $\chi^2$ for $n$ data points and $n_p$ parameters: 

\begin{align}
    \epsilon &= {Y} - {A}{X}\\
    \chi^2 &= \frac{\epsilon^T W \epsilon}{n - n_p} \\ 
    (A^T W A)^{-1} \chi^2 &=     \begin{bmatrix}
    \left(\sigma^{fit}_{{log(T)}}\right)^2 & {\rho_{12}} \\  
    {\rho_{21}} & \left(\sigma^{fit}_{p}\right)^2 \\ \label{eq:Covariance}
    \end{bmatrix} 
\end{align}

The 1-$\sigma$ uncertainty in our parameters corresponds to the square root of the diagonal elements of the covariance-variance matrix in \Cref{eq:Covariance}, {while the off-diagonal elements $\rho$ are the correlation between the factors. From there,} we obtain the uncertainties for the brightness temperature $\sigma^{fit}_T$ (by taking the exponential of $\sigma^{fit}_{{log(T)}}$) and for the limb-darkening parameter $\sigma^{fit}_p$. {In \Cref{fig:FitUncertainty} we verify the fit uncertainties derived from the weighted least square approach through the use of a computational more expensive Markov-Chain Monte-Carlo (MCMC) approach, which samples the solution space and the correlation between the parameters based on the same parameterization in \Cref{eq:Tld}. 
The best-fit quantities derived from each other agree within the derived uncertainties, with the MCMC finding larger uncertainties for the derived values. We therefore will carefully assess the impact of the uncertainties on the derived ammonia distributions. The MCMC results also indicate the correlation between the two quantities, where an increase in temperature correlates well with an increase in brightness temperature. We note however that the absolute calibration uncertainty for temperature (1.5\% $\approx$ 5K corresponding to the example in \Cref{fig:FitUncertainty}) is an order of magnitude larger than the fit uncertainty(0.15K), and as such the correlation is much smaller compared to the absolute calibration. Therefore, we treat them as independent errors in our analysis.}
The final uncertainty for the brightness temperature is a combination of the fit uncertainty $\sigma_{T}^{fit}$ and the absolute calibration uncertainty $\sigma_{T}^{cal}$ \citep{Janssen2017}, which we add in quadrature. We adopt an absolute calibration uncertainty $\sigma^{cal}_T$ of 10\% for Channel 1 due to the unknown opacitiy of absorbers (e.g., ammonia, water, and alkalis) at such high pressures \citep{Hanley2009, Bellotti2016, Bellotti2017,Bellotti2018}, and 1.5\% for all other channels \citep{Janssen2017}. The final uncertainty for the limb-darkening coefficient is a relative measurement and therefore corresponds only to the fit uncertainty.

\begin{align}
\label{eq:sigmaT}
    \sigma_T &= \sqrt{(\mathbf{n}\; \sigma_{T}^{fit})^2 + (\sigma_{T}^{cal})^2}\\ 
    \sigma_p &=\; \;  \mathbf{n} \; \sigma_{p}^{fit}
\end{align}

\begin{figure*}[h]
\centering
\includegraphics[width=0.8\textwidth]{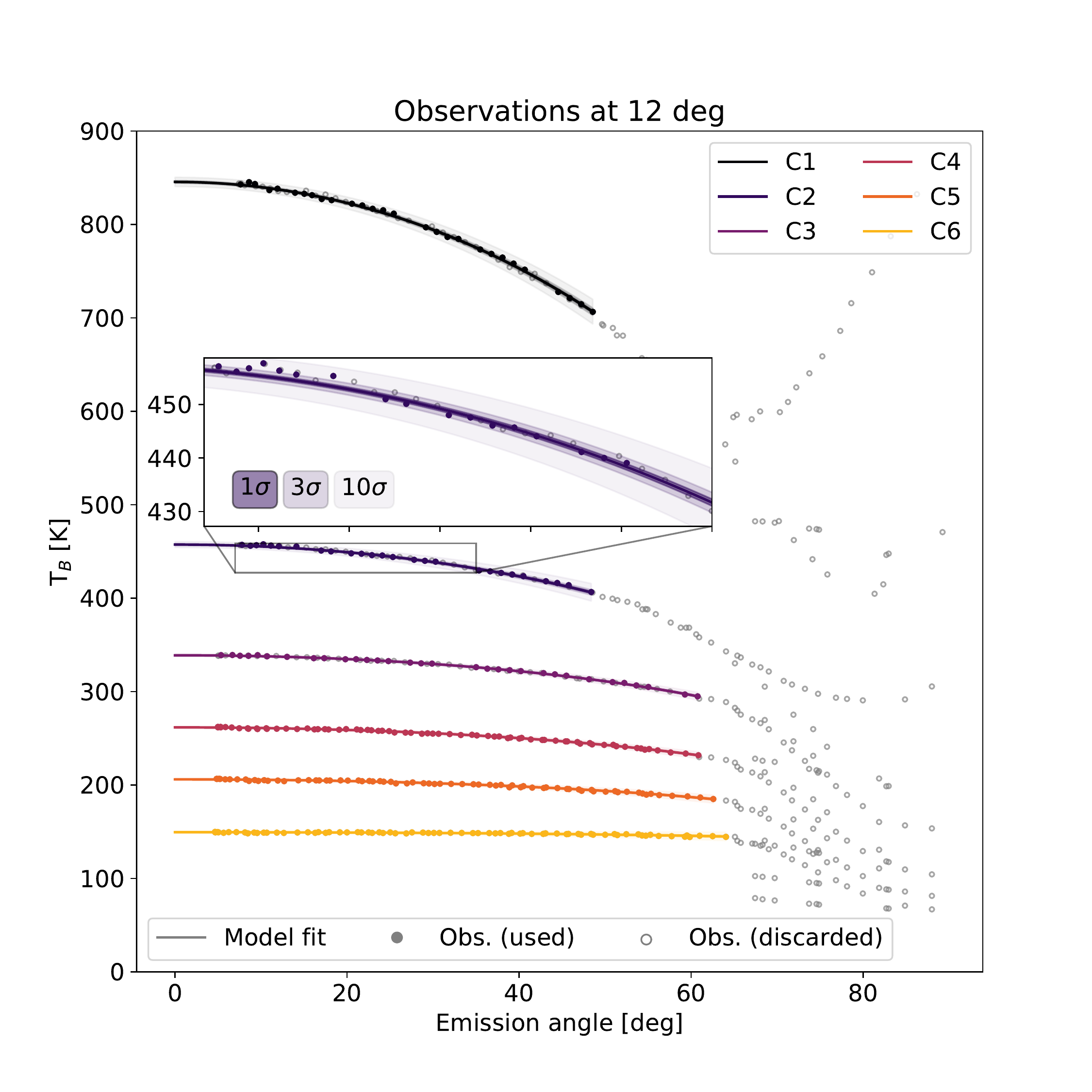}
\caption{ Brightness temperature at 12$^{\circ}$ deg north latitude for the 6 different channels, plotted against the beam-convolved emission angle. The closed circles correspond to the observations considered in the fit, and the open circles to the observations that were removed. Open circles inwards of the cut-off are observations with high susceptibility to synchrotron radiation (spatial filter -- see Section \ref{ssec:ds}), while open circle outwards are measurements where part of the beam is off the planet. The cut-off is determined such that all beam elements interior 2.5 HPBW are on the planet.  The impact of the synchrotron radiation leaking in through the side lobes can be seen in Channel 1 and 2, where the brightness temperature increases beyond 60 deg, while the drop in brightness temperature at higher frequencies is due the cold sky leaking in. The solid line with shading corresponds to the fitted model according to \Cref{eq:Tld} and three uncertainty fidelities (n=1,3,10). The intersect at an emission angle of zero is the nadir brightness temperature, and the limb-darkening parameter defines the shape of the curve. }
\label{fig:TBandLDfit}
\end{figure*}

\begin{figure*}[h]
\centering
\includegraphics[width=0.6\textwidth]{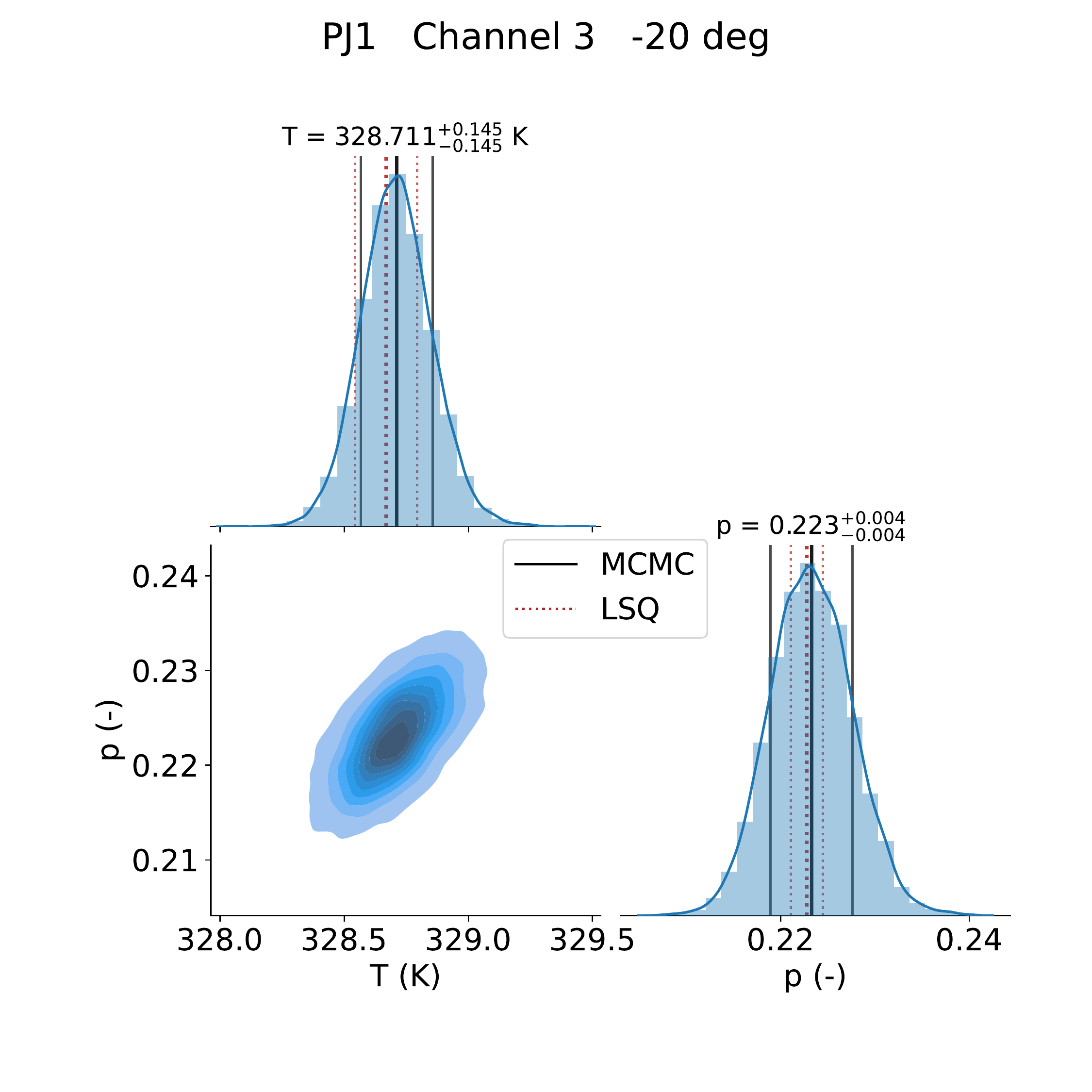}
\caption{{Verification of the fit uncertainty derived from the weighted least squares approach (see \Cref{eq:wlls}) by comparing the results to an Markov-Chain Monte-Carlo (MCMC) method using the same underlying model to fit the data (see \Cref{eq:Tld}). The solid black lines corresponds to the 16\%, 50\% and 84\% quantiles, while the dotted red lines indicate the errors retrieved from the weighted least square approach. The best-fit quantities for both methods agree well within their uncertainties. The largest difference is found in the uncertainty of the fitted limb-darkening parameter, where the weighted least square errors are smaller than the MCMC results.}}
\label{fig:FitUncertainty}
\end{figure*}

Spatial variations {of the parameters} within the beam and remaining synchrotron contribution both deteriorate the reliability of the fit, making the retrieval prone to errors in case of an overreliance on these measurements. To fully understand the impact of the uncertainties, we allow $\mathbf{n}$ to vary between 1 and 10 (inset in \Cref{fig:TBandLDfit} shows the various uncertainty levels), and test the robustness of our results to the uncertainty fidelity $\mathbf{n}$.  {If our models are sufficiently capture the complex atmospheric structure, the choice of uncertainty should not matter.} Thus, as a function of latitude, we have reduced the MWR measurements to two parameters: nadir brightness temperature $T_{b0}$ and limb-darkening coefficient $p$ (with a 1 deg step-size in latitude per PJ) and their corresponding uncertainties. \\ 

In addition to the individual PJs, our final goal is to produce a global (i.e., longitude, or orbit-averaged) latitudinal distribution of trace gases. We therefore combine the individual PJs and compute the mean for each channel. We caution the reader that the global mean is based on nine narrow slices of the atmosphere separated both in time and longitude, and is thus not a perfect representation of a global atmosphere. For the uncertainties, we elect not to use the standard deviation between the individual PJs since they (i) reflect spatial variations instead of measurement uncertainties and (ii) do not reflect the absolute uncertainty for the temperature measurements. Instead we compute the average fit uncertainty $\overline{\sigma}^{fit}$ at a given latitude among the PJs and add them to the absolute uncertainty: 

\begin{align}\label{eq:sigmafit}
    \sigma_T^{mean}  &= \sqrt{ \left(\mathbf{n}\;\overline{\sigma}_T^{fit}\right)^2 +\left(\sigma_T^{cal}\right)^2}\\
     \sigma_p^{mean} &=\;\; \mathbf{n}\;\overline{\sigma}_p^{fit}
\end{align}

We use $\mathbf{n}$ = 3 (i.e., 3-$\sigma$) for the fit uncertainty when plotting the trace gas distributions, but we also show the retrieved atmospheric quantities for different fidelities (n=1,5,10) in Section \ref{sec:results}.

\subsection{Data selection} \label{ssec:ds}
Despite limiting observations to viewing geometries with 99\% of emissions coming from the planet, the low frequency observations show signs of synchrotron contamination. The tell tale sign of synchrotron contamination is a breakdown of the typical relationship between emission angle and brightness temperature. Observations for which the synchrotron radiation off the planet is stronger than the atmospheric signal show a non-physical limb-brightening effect, unlike the limb-darkening that is apparent for atmospheric observations. In the absence of accurate synchrotron radiation models, and given the complexity of the electron distribution and the magnetic field close to the planet \citep[e.g., ][]{dePater1981,dePater1997,SantosCosta2017,Moore2019}, we use two different filters to remove observations that are suspect to contamination. 

\subsubsection*{Spatial Filter}
The main synchrotron radiation belt of Jupiter is constrained to low latitudes around the magnetic equator \citep[e.g., ][]{roberts1976pitch,dePater1981,sault1997first,dePater1998}. Despite the low sensitivity of the antenna beam in the side-lobes, the combination of high intensity radiation and large emission area causes problems in observations where synchrotron emission enters through the side lobes. We therefore restrict measurements where the sensitivity to the synchrotron radiation in the side lobes is minimal. In practice, when the spacecraft is at higher latitudes (10-60$^{\circ}$), we remove all observations that are looking towards the equator (where the equatorial synchrotron radiation is leaking in) and instead use only observations that look polewards. See \Cref{fig:TBandLDfit} and Appendix B for examples of this filter. 

\subsubsection*{Limb-Darkening Filter}
The fitted limb-darkening coefficient and its deviation from the mean is another constraint that we use to assess the synchrotron contribution to the observations. Observations that are bright at the limb (i.e., a lower than expected value for the limb-darkening) are a telltale sign of contribution by synchrotron radiation. Latitudes where the limb-darkening coefficient deviates more than 5 standard deviations from the mean for the given orbit are replaced by the orbit mean at that latitude.

\subsection{Orbit average} 
\begin{figure*}[h]
\centering
\includegraphics[width=\textwidth]{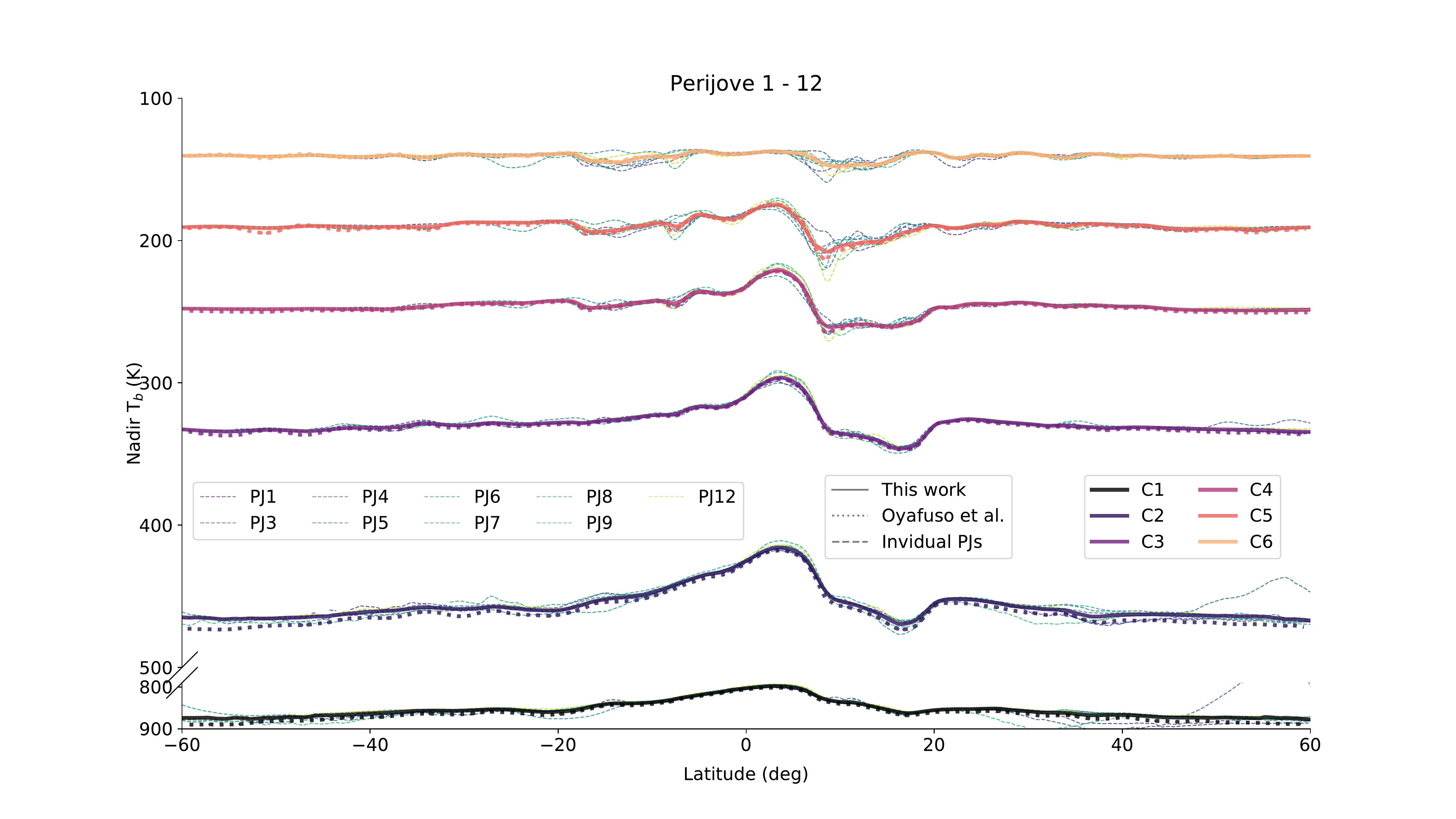}
\caption{Nadir brightness temperature for the 6 channels between -60 and 60 deg, where the thick lines indicate the orbit averaged values, and the dashed thin lines correspond to the individual PJs. The dotted line represents the orbit averaged value based upon a deconvolution algorithm  \protect{\citep{Oyafuso2020}} and shows a good agreement between the two methods, validating both our approaches. }
\label{fig:Juno_average_T}
\end{figure*}

\begin{figure*}[h]
\centering
\includegraphics[width=\textwidth]{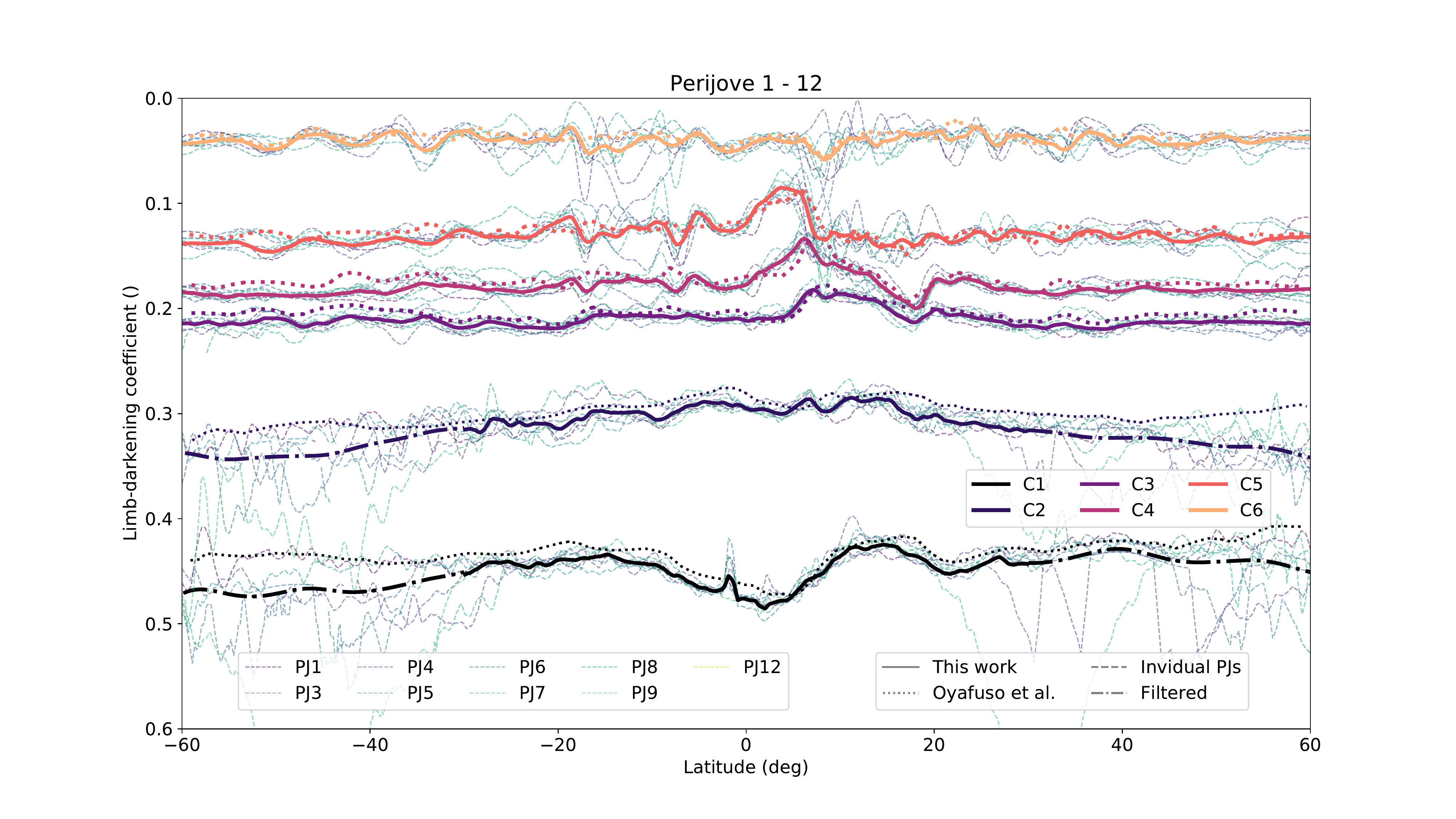}
\caption{Same as \Cref{fig:Juno_average_T} but for the limb-darkening parameter.}
\label{fig:Juno_average_p}
\end{figure*}

The geometry during each flyby is dictated by the type of science that is prioritized. For MWR observations, we therefore restrict ourselves to a subset of nine orbits from the first 12 to compute the orbit average (see \Cref{tab:JunoCampagain}). We use the retrieved uncertainties in the parameters as weights to compute the mean profiles for brightness temperature and limb-darkening after applying the spatial and limb-darkening filters discussed above. \\ 
The results of the data reduction for these Juno data are shown in Figures \ref{fig:Juno_average_T} (nadir brightness temperature) and \ref{fig:Juno_average_p} (limb-darkening coefficient), where the solid lines correspond to the mean and the thin dashed lines correspond to the various PJ passes. For verification purposes, we also compare our results to published orbit-averaged data of \cite{Oyafuso2020}, indicated by the dotted line. The variations with latitude of radio-cold and radio-warm regions indicate the presence of enriched and depleted trace gases, which modify the opacity for the escaping blackbody radiation {, in line with our assumption that the variations in kinetic temperature are small}. Some of the latitudinal variations seen in the mean profiles can be traced down across all channels, such as the radio cold region around the equator, indicating the presence of atmospheric variations spanning several orders of magnitude of pressure. At higher latitudes, the profiles show that the variations decay rapidly with depth, highlighting more shallow atmospheric features.  Atmospheric variability, as seen by the dispersion between individual PJ profiles, is seen especially at the higher frequencies, which correspond to the upper layers of the atmosphere where temporal and spatial variability dominate \citep{dePater2016}. \\ 
The structure of the limb-darkening is more chaotic, but to first degree it mirrors a structure similar to the brightness temperature, with the biggest diversions seen in the equatorial zone and the neighboring north equatorial belt. We can see smaller features at higher latitudes corresponding to regions where the white zones and brown-reddish belts alternate. 
The limb-darkening coefficient is very sensitive to variations in spatial structure within the binned regions, so that it becomes difficult to interpret the signal at higher frequencies, and we caution reading too much into the fine scale structure at the highest frequencies. 

\section{Methods} \label{sec:methods} 
The presence of radio-absorbing trace gases in a planet's atmosphere modulates the thermal emission from the atmosphere as it propagates upwards. By prescribing a trace gas distribution with a {pre-defined} temperature profile for the atmosphere of Jupiter, we can simulate its brightness temperature using a radiative transfer (RT) model, such as that described by e.g., \citet{Berge1976, dePater1985}. Such simulated data (see Section \ref{sec:RT} for the RT code) can then be compared to Juno MWR and VLA observations: by adjusting the abundances and altitude profiles of radio-absorbing trace gases (and/or the temperature profile \citep{Fletcher2021}), one can find a best fit to the data. In Section \ref{sec:results} we focus on the $NH_3$ abundance, which is the dominant absorber at the frequencies Juno MWR and the VLA are operating \citep[e.g., ][]{dePater1986}. At lower frequencies the presence of water (vapor and cloud particles) starts to have an effect on the observations as well \citep{dePater2005}. This signal, however, is difficult to interpret and we test the impact of the cloud particles on the retrieval in Section \ref{ssec:ezinv}. 

But what is the best way to prescribe the distribution of trace gases? In the past, researchers determined ``arbitrary'' jumps in the ammonia profile \citep[e.g., ][]{dePater1986} to find a best fit to their data. \citet{dePater2001} pointed out that profiles which gave a best fit to the available data did not follow the profile expected from thermo-chemical equilibrium calculations \citep{Weidenschilling1973,Atreya1985}. \citet{Li2017} followed a similar route, opting to split the atmosphere in several layers and then use a Markov-Chain Monte-Carlo (MCMC) algorithm to find the stochastic best fit to Juno PJ1 observations. Considering the few Juno observations that we have (i.e., data at 6 frequencies), the problem is under-constrained, that is, we have more variables to fit than observations constraining them. Additionally, the received radiation is an integrated signal over many layers of the atmosphere, and as such the inversion is not unique. {The problem can be further constrained}  through regularization of the atmosphere, limiting how much the atmosphere can change between {pressure levels} \citep{Li2017}, or through the use of thermo-chemical equilibrium models as mentioned above \citep{dePater2016, dePater2019}. We know from observations of the troposphere on Earth that the atmosphere is more complex than a simple thermo-chemical equilibrium model prescribes. Nevertheless, it adds {a-priori information}, and we opt to modify the thermo-chemical equilibrium models by introducing factors that quantify the departure from equilibrium in these models. 

\subsection{Model 1: Perturbed thermo-chemical equilibrium model}\label{ssec:TEmodel} 
In the absence of other observations, the simplest description of an atmosphere is a well-mixed atmosphere, where the trace gas (\footnote{We express trace gases in both parts per million (ppm) and in proto-solar abundances: $\frac{C}{H^{2}}$, = 5.9e-4, $\frac{N}{H^{2}}$ = 1.48e-4, $\frac{S}{H^2}$ = 2.89e-5, $\frac{O}{H^2}$ = 1.07e-3 \citep{Asplund2009}}) abundance ($x_i$) is constant with pressure until the saturation pressure, $P_{sat}$, at temperature $T_c$ for the specific trace gas is reached, at which point the gas starts to condense out to form clouds. We modify the relationship by introducing the relative humidity $r_h$ in \Cref{eq:rhnh3}, where a value greater than unity indicates super-saturation, delaying the condensation of cloud particles. On Earth, super-saturation occurs due to a variety of reasons, most commonly the absence of cloud condensation nuclei, or the presence of strong updrafts. Sub-saturation at these levels, corresponding to values less than unity, are most likely a reflection of averaging areas inside the radio beam that are saturated and areas that are not saturated.  

\begin{equation}\label{eq:rhnh3}
 NH_3(P) = r_h \frac{P_{sat}(T)}{P} 
\end{equation}

Below the ammonia ice clouds, the absence (or lack of detection) of hydrogen sulfide (H$_2$S) gas on Jupiter has been explained by the formation of ammonia hydrosulfide clouds (NH$_4$SH), where essentially all the H$_2$S is trapped in the NH$_4$SH cloud as it combines with the more abundant NH$_3$. The onset of cloud formation starts when the product of the two partial pressure controlled by the partial pressure of the two gases, and the reaction rate has been parameterized by \citet{Weidenschilling1973} and \citet{Atreya1985}. 

\begin{equation} 
 NH_3 + H_2S \rightleftharpoons  NH_4SH
\end{equation} 

The only measurement of sulfur on the planet so far comes from the Galileo probe measurements \citep{Niemann1996,Wong2004} in a so-called 5-$\mu$m hot spot (i.e., a region devoid of clouds), which showed a solar enrichment of H$_2$S similar to that of NH$_3$. Since our results for the ammonia abundance at high pressures ($\gtrsim$20 bar) are lower than the estimates from the Galileo experiment (see \Cref{fig:DA}), we opt to not use the Galileo probe measurement. {Instead we retrieve the H2S abundance indirectly through its effect on the NH4SH cloud formation, which equally removes NH$_3$ from the atmosphere. Additionally, we} account for uncertainties in the pressure at which the NH$_4$SH cloud forms by introducing a factor, $\eta_{NH_4SH}$, to the equation by \citep{Lewis1969} that allows the cloud formation to be delayed.  At a value of one, the reaction occurs at the predicted pressure, and for values less than one, the clouds form higher up in the atmosphere than predicted by thermo-chemical equilibrium models: 

\begin{equation} 
    log_{10}(P_{NH_3} P_{H2S}) = \eta_{NH_4SH} log_{10} K = \eta_{NH_4SH} (-4715/T + 14.83 ) 
\end{equation} 

We did not find a constant ammonia abundance below the NH$_4$SH cloud layer that could explain both the VLA and Juno observations; therefore, we introduced a parameterization for the region between an assumed deep, globally uniform atmosphere and the NH$_4$SH cloud layers. Processes such as dynamics \citep{Showman2005} or microphysics \citep{Guillot2020} can locally enrich or deplete the atmosphere, leading to gradients in abundance with pressure. We started with the simplest model by fitting a single mixing parameter between a top and a bottom pressure, and a deep ammonia abundance. The mixing gradient is defined such that a  positive gradient constitutes an increase in ammonia with increasing pressure. 

\begin{align}
     NH_3(P) &= NH_3^{deep} - H_{mix} (ln(P_b) - ln(P)) \\\label{eq:dnh3dp}
     H_{mix} &= \frac{\delta NH_3}{\delta ln(P)}
\end{align}


We postulate that at some depth, the atmosphere is well mixed across all latitudes. The most well-suited observations for constraining a global deep abundance come from MWR Channel 1, which probes deepest into the atmosphere (see \Cref{fig:WF}). Unfortunately, the inversion of the observations at low frequencies is also the most uncertain, due to scarcity of high pressure and temperature experiments simulating the conditions in the gas giants \citep{Hanley2009,Bellotti2016,Bellotti2017}. Different published ammonia absorption coefficients lead to differences as large as {many tens of Kelvin for modeled temperature in Channel 1}. We therefore opt to reduce the influence of the lowest frequencies, by applying large absolute uncertainties to the Channel 1 measurements, so that we instead rely on the higher frequency channels. Lastly, for the global ammonia abundance maps, we opt to prescribe the deep ammonia abundance and only derive the pressure below which the atmosphere is well mixed. This is in line with our assumption that there exists a deep, well-mixed reservoir of trace gases on the planet.

While H$_2$S gas on Jupiter does not contribute noticeably to the absorption at radio wavelengths, water has a non-negligible effect on the observations at frequencies below $\sim$6 GHz \citep[see, e.g., ][] {dePater1985,dePater2005,Janssen2005}. For the main analysis, we enriched the water abundance by the same solar enrichment factor that we retrieved for the ammonia gas (2.30), and let it condense out at the water condensation point. This is the simplest assumption and is in line with the results for the Equatorial Zone (EZ) \citep{Li2020}. For this part of the analysis, we included only an indirect effect of the water cloud on the ammonia: about 3\% of gaseous ammonia can be absorbed into the liquid water rain drops \citep{Weidenschilling1973}. This effect causes a small local decrease in ammonia at the water cloud level, $\sim$6 bar.

\subsection{Model 2: Stochastic variations between pressure nodes} \label{ssec:Smodel}
{In addition to the perturbed thermo-chemical equilibrium model, we developed a model similar to the one used by \cite{Li2017}, that fits gradients between pressure levels. Moving up from 100 bar we introduce pressure nodes between which we assume an exponential gradient that mimic processes locally enriching or depleting the atmosphere in trace gases. We set the pressure nodes to 40, 20, 15, 10, 6, 4, 2 and 1 bar above which the atmosphere follows the saturation-vapor pressure curve as defined in \Cref{eq:rhnh3}. Since the problem is underconstrained we introduce a 'lasso'-regularization constraint \citep{tibshirani1996} that aims at smoothing the vertical variations, penalizing large gradients between pressure nodes (see \Cref{eq:C2}).}

\section{Radiative Transfer Model}\label{sec:RT} 
We use our open-source radiative transfer model radioBEAR (Radio BErkeley Atmospheric Radiative-transfer\footnote{https://github.com/david-deboer/radiobear}) to simulate the brightness temperature for a given distribution of trace gases and a prescribed temperature structure in the atmosphere by splitting the atmosphere into multiple layers and calculating the contribution from each individual layer. This model has been described by \citet{dePater2005,dePater2014,dePater2019}. The opacities for ammonia and water were derived from lab experiments that simulate the high-pressure environment for a given hydrogen-helium atmosphere by \cite{Devaraj2014,Hanley2009,Bellotti2016, Bellotti2017}. For the Channel 1 frequency, the \cite{Bellotti2016,Bellotti2017} modeled brightness temperatures were too high and not reconcilable with the Juno observations, so that we favored the earlier lab measurements as did \citet{Li2017}.

\subsection{Temperature Structure}\label{ssec:TPprofile}
The temperature structure in the atmosphere is a crucial element for the retrieval, since both trace gas and temperature variations can alter the observed brightness temperature, and we can only observe the combined effect. While observers have used mid-infrared wavelengths to derive the temperature at pressures $P \lesssim 0.7$ bar \citep{Fletcher2009}, the temperature structure below the radiative-convective boundary cannot be obtained with these techniques. 

The temperature at the depth where radio waves originate are controlled by atmospheric dynamics, microphysics, and wave patterns in the atmosphere, and to date, there are very few temperature measurements below the radiative-convective boundary ($\sim$0.5 bar). The first measurements come from the Voyager flyby, which used radio-occultation measurements to probe the temperature structure down to the 1 bar level on ingress and egress to measure the temperature of the North Equatorial Belt (NEB, 165 $\pm$ 5 K) and the Equatorial Zone (EZ, 170.4 $\pm$ 5 K), respectively \citep{lindal1981atmosphere,Guillot2004}. The latent heat release of condensation in the EZ decreases the temperature gradient in the atmosphere, leading to higher temperatures at the same pressure levels compared to an atmosphere that follows a dry adiabat. Hence, the observed temperatures in the EZ and NEB are consistent with a local enrichment of trace gases in the EZ compared to the NEB. The only true in-situ measurements come from the Galileo probe experiment, which entered a highly trace-gas depleted region of the atmosphere, somewhat similar to the conditions found in the NEB, and measured a temperature structure resembling a dry adiabat down to 20 bars. 

\begin{figure*}[h]
\centering
\includegraphics[width=0.75\textwidth]{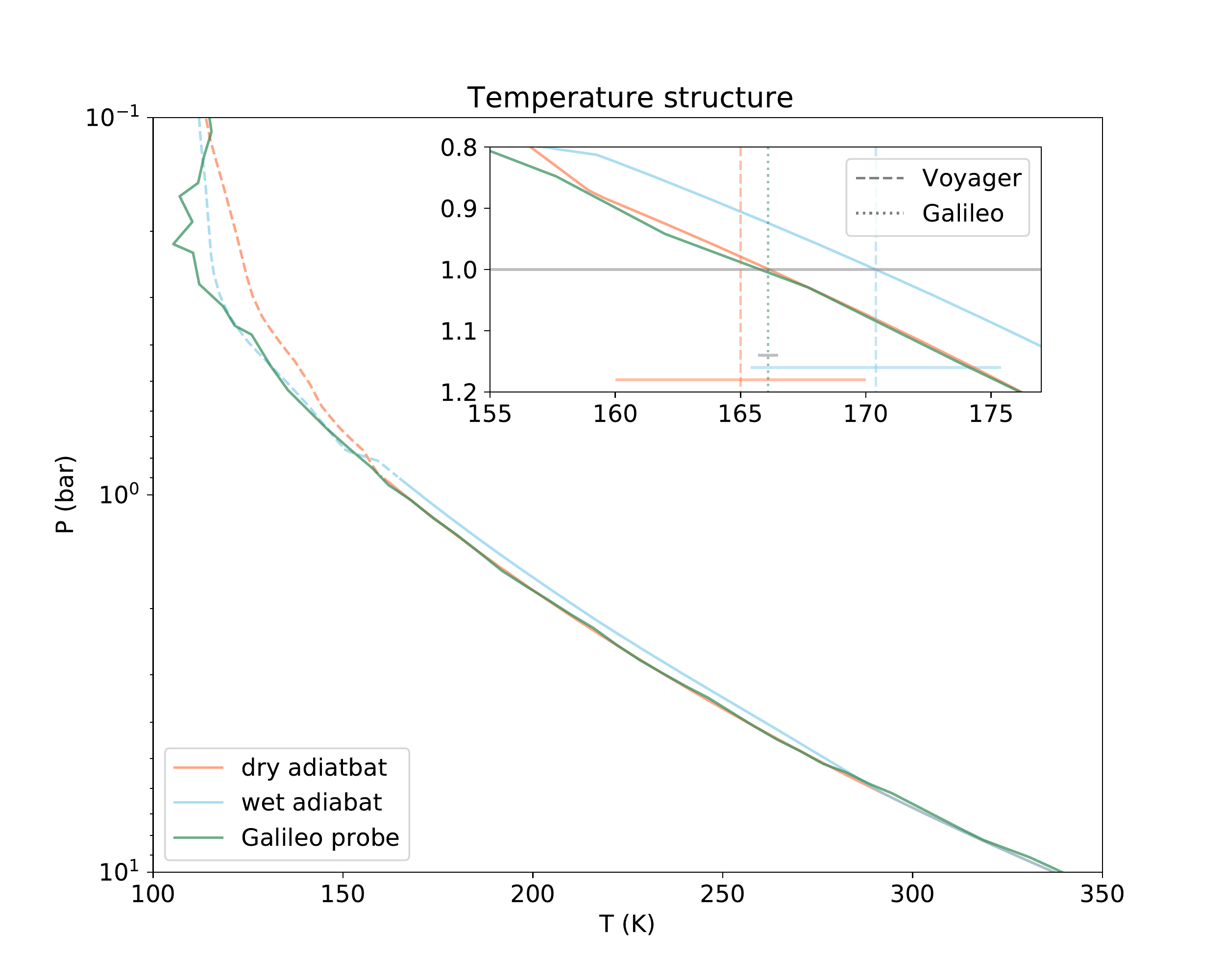}
\caption{Compilation of the  relevant temperature profiles which we assume for our retrieval. The temperature profiles are anchored at the 1 bar level to the radio occultation measurements of Voyager \citep{lindal1981atmosphere} and to the Galileo probe measurements \citep{Seiff1996}. Above the 0.7 bar level we use profiles obtained via infrared measurements \citep{Fletcher2009}. Below the 1.0 bar level we use the expected wet and dry adiabatic lapse rate until the two profiles merge at the height of the water cloud. The little cutout shows the anchor points at the 1 bar level, with the corresponding uncertainties.}
\label{fig:TPprofile}
\end{figure*}

We have built two temperature profiles---a wet and a dry adiabat---that are anchored to the Voyager and Galileo measurements (\Cref{fig:TPprofile}). Starting from deep within the planet, the absence of condensible species implies that the temperature follows a dry adiabat: 

\begin{equation}
    \Gamma_d = \frac{dT}{dz} = \frac{g}{\sum_i \frac{C_{p_i}}{x_i}}
\end{equation}

{The heat capacity $C_p$ of an parcel of air is the sum of the heat capacities of the individual atmospheric species weighted by their mixing ratio (see \citep{dePater1985}). We updated the heat capacity for normal, para- and ortho-hydrogen based on the formalism of \citet{Leachman2009} that is accurate to within 0.1\% compared to experimental results until 1000K, corresponding to about 400bar. For the gravity, we take the gravity coefficients up until $J_4$ into account to model the gravity as a function of latitude and depth \citep{Iess2018}. The profile follows the dry adiabat} until the water condensation level is reached, at which point the two temperature profiles start to diverge. While the dry adiabat reaches 166.1 K at the 1 bar level (Galileo anchor point \citep{Seiff1996}), the latent heat from water decreases the temperature gradient for the wet adiabat, so that in the EZ a temperature of 170.4 K is reached at the 1 bar level. Higher up in the atmosphere (P $\lesssim$ 0.7 bar), we use profiles obtained from the mid-infrared \citep{Fletcher2009} and merge the two temperature profiles to cover the full depth which radio waves probe. In the radiative zone the profiles reverse, that is, the NEB becomes warmer than the EZ as shown by the dotted lines in \Cref{fig:TPprofile}.

{For our global ammonia maps we prescribe a dry adiabat, however, we test the impact of a wet adiabat on the vertical ammonia distribution in Section \ref{ssec:ezinv}}. 

\subsection{Weighting functions} 
\begin{figure*}[h]
\centering
\includegraphics[width=\textwidth]{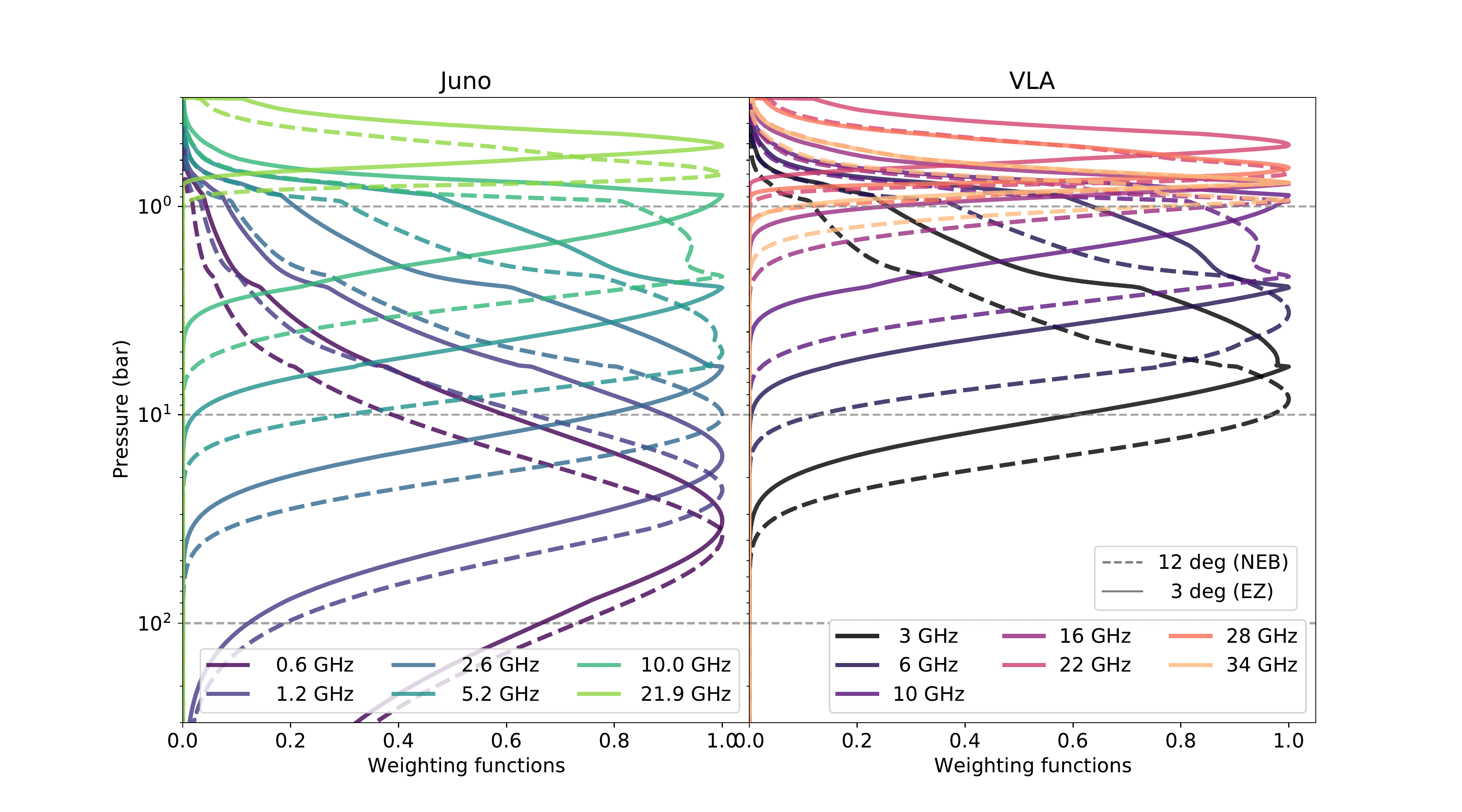}
\caption{Weigthing functions visualize the pressure levels from which radiation is received. We plot the functions for both end members on the planet, a highly enriched part in the atmosphere (solid line corresponding to the EZ), and a highly depleted part (dashed line corresponding to the NEB) showing the depth and range from which we receive the signal. The abundance profile for which we compute the weighting functions are taken from our fit to the orbit-averaged MWR data (see \Cref{fig:NH3-PJmean}). The left hand side shows the relevant functions for the Juno spacecraft, and the right hand side shows the function for the VLA. }
\label{fig:WF}
\end{figure*}

Thermal emission captured at radio wavelengths originates over a range of atmospheric pressures, and the final signal is the integrated emission over these layers. The weighting functions illustrate this effect numerically, by showing the depth of and range from where we receive the emission. We plot the weighting functions in \Cref{fig:WF} for the EZ and the NEB, exemplifying the end members in terms of ammonia gas distribution covering the most enriched and most depleted region of the planet. At lower frequencies, the radio waves originate over a wide range of pressures going down to a few hundred bars, while at higher frequencies the weighting functions become more peaked, indicating that they probe a smaller pressure range. The center of the absorption feature is at 22 GHz, so that observations at this frequency probe highest in the atmosphere, while observations both at higher and lower frequencies probe deeper into the atmosphere. Lastly, the depletion of trace gases in the NEB allows for probing deeper into the atmosphere, enabling retrieval of ammonia abundances at higher pressures.  

\subsection{Optimization} 
For a given distribution of trace gases, we can simulate the brightness temperature and the limb-darkening coefficient; these modeled parameters are referred to as $T_m$ and $p_m$. By matching our simulated atmosphere to the observations, we retrieve the best-fit distribution of ammonia gas. We opted to retrieve the distribution per 1-deg bins, independent of neighboring latitudes, for a robust retrieval. We first determined the optimum using the scipy \footnote{https://www.scipy.org/} Nelder-Mead optimizer to find an atmospheric composition that minimized the difference between the model ($T_m$, $p_m$) and the observations ($T_{b0}$, $p$), weighted by the uncertainties ($\sigma_{T_{b0}}$, $\sigma_p$) as shown in \Cref{eq:C1}. {It is here that the choice of 'n' for the uncertainties has a largest impact on where the optimal solution is, as the cost function is modulated by the uncertainties.} The uncertainties are essential for combining the various observational quantities and weighing them appropriately. 

\begin{equation}\label{eq:C1}
    C_{Model 1} = \sum_{1}^{6} \frac{(T_{m} - T_{b0})^{2}}{\sigma_{T_{b0}}^{2}} +  \sum_{2}^{5} \frac{(p_{m} - p)^{2}}{\sigma_{p}^{2}}
\end{equation}

For our stochastic model we include a regularization function that penalizes large gradients ($ \frac{\delta NH_3}{\delta P}$) between pressure nodes. We chose $\lambda = 1E4$\footnote{The mixing gradient in this paper is always converted to parts per million for ease of understanding} with which the penalty is much smaller in magnitude than a typical cost function result, to avoid interferences with the retrieval. 
\begin{equation}\label{eq:C2}
    C_{Model 2} = C_{Model 1} + \lambda \left|\sum{ \frac{\delta NH_3}{\delta ln(P)}_i}\right|
\end{equation}

We only use the limb-darkening information from Channel 2 to Channel 5. We discarded Channel 1 limb-darkening measurements as we cannot match the opacity of Channel 1 with only ammonia. Other absorbers such as water \citep[(both vapor and droplets)]{dePater2005} and alkalis \citep{Bellotti2018} are known to add opacity at these low frequencies. We also discarded Channel 6 because it is very insensitive to our variations in the ammonia abundance and mostly driven by spatial variations (see the VLA maps published in \citet{dePater2016,dePater2019}). This established a combination of parameters that minimize the difference, but does not indicate the range of possible solutions within the constraint. We therefore coupled our retrieval with a Markov-Chain Monte-Carlo (MCMC) \citep{Foreman-Mackey2013} for the set of parameters $x$ to find the uncertainties in the fit parameters and the correlations between the parameters. The MCMC maximizes the log-likelihood function $ln p$: 

\begin{equation}
    ln p(T,p|x,\sigma) = -0.5\left(\sum_{1}^{6}  ln(2\pi\sigma_{T}) + \frac{(T_{m} - T_b)^{2}}{\sigma_T^{2}} +  \sum_{2}^{5}  ln(2\pi\sigma_p) \frac{(p_{m} - p)^{2}}{\sigma_{p}^{2}} \right)
\end{equation}

\section{Results} \label{sec:results} 
In this section we present our best-fit ammonia distribution for the average of the first 12 orbits when MWR was able to observe the atmosphere (see \Cref{tab:JunoCampagain}) and compare our results for the first PJ to the inversion by \citet{Li2017}. 

\subsection{Well-mixed reservoir} \label{ssec:DA}
The first quantity we constrain is the amount of ammonia in the region that is globally well-mixed. Based on \Cref{fig:WF}, we can see that even at the lowest frequency observations of Juno, the signal from depths over 50 bar is poorly constrained, due to the wide range of pressures the signal originates from. There are four observed quantities from the orbit-averaged quantities that constrain the deep ammonia abundance: the brightness temperatures of Channel 1 to Channel 3, and the limb-darkening coefficient of Channel 2. Since our model cannot find an ammonia distribution that fits all four quantities, we instead search for a solution that minimizes the overall error. The key parameter we vary is the factor $n$ for the fit uncertainties ($T_b \pm n*\sigma^{fit}_{T}$,$p \pm n*\sigma^{fit}_{p}$) and we test the sensitivity of our solution to the chosen fidelity factor $n$. As a reminder, a value of 1 corresponds to the 1-sigma results when fitting the nadir brightness temperature and limb-darkening (\Cref{eq:sigmafit}). We consider four different combinations: 

\begin{itemize} 
\item[1-$\sigma$] This option uses the mean error in the \textbf{fit uncertainty} based on the nine limb-darkening profiles. This combination maximizes the fit to the limb-darkening coefficient of Channel 2, at the cost of fitting the brightness temperature in Channel 1 and Channel 2. We interpret this number as a lower limit to the deep abundance. 
\item[3-$\sigma$] This combination compromises the fit between all quantities, and we use this for the final result. 
\item[5-$\sigma$] Similar to 3-$\sigma$, where the limb-darkening coefficient still influences the retrieval. 
\item[10-$\sigma$] With these large uncertainties, the brightness temperature is the main quantity driving the retrieval. While the limb-darkening uncertainty changes by a factor of 10, the total uncertainty in the temperature changes by less than 2\% due to the large calibration errors. This retrieval we interpret as an upper limit to the retrieved ammonia abundance. 
\end{itemize}

\begin{figure*}[h]
\centering
\includegraphics[width=0.75\textwidth]{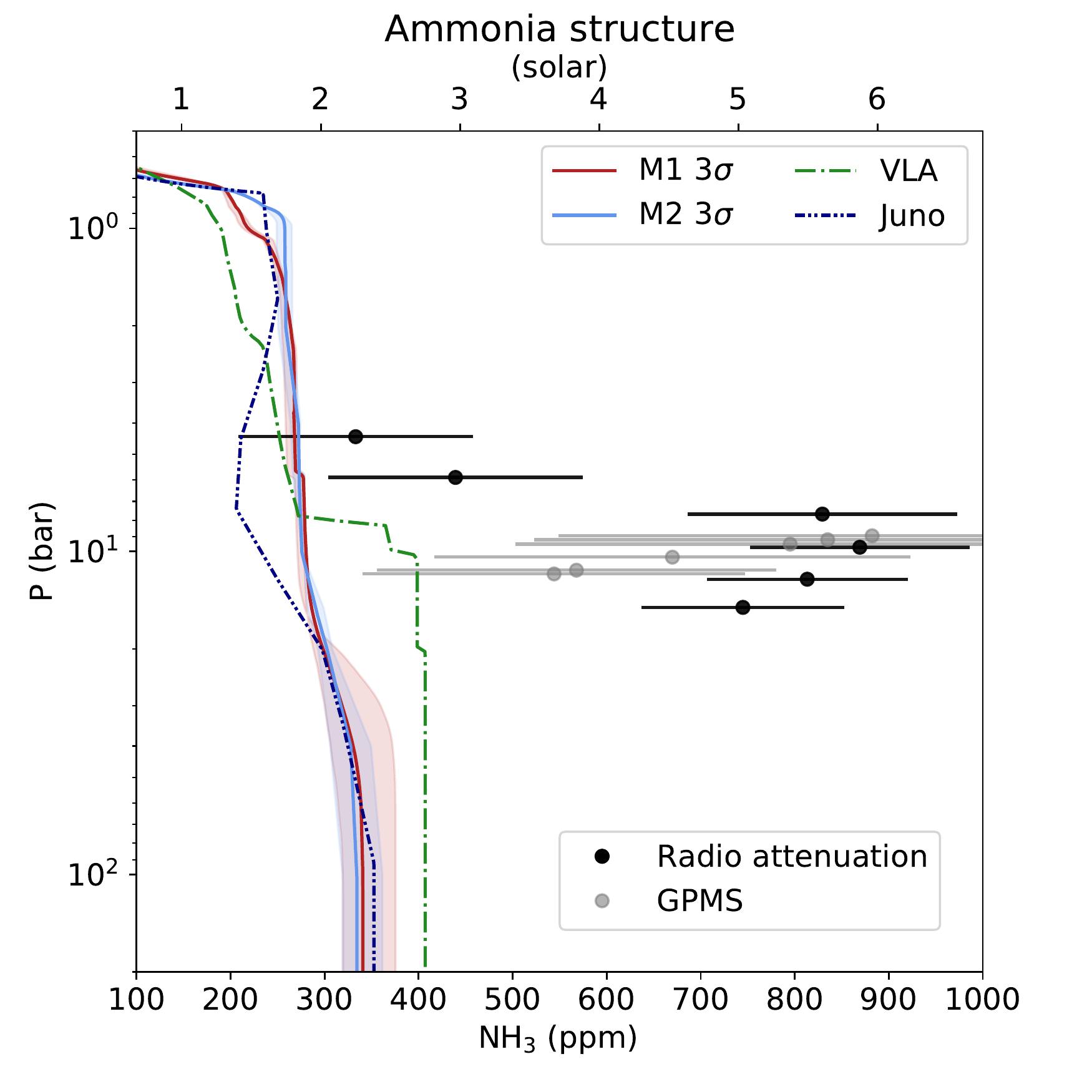}
\caption{Globally averaged [-45$^{\circ}$,45$^{\circ}$] ammonia profile as a function of depth. The red profiles corresponds to our thermo-chemical equilibrium model, and the blue line corresponds to our stochastic model with the shading corresponding to n = 1,10. The blue dashed curve is based on the results  of \citet{Li2017}, and the green curve is based on the VLA data from \citet{dePater2019}. The radio attenuation \citep{Folkner1998}  and Galileo Probe Mass Spectrometer (GPMS)\citep{Wong2004} are superimposed. }
\label{fig:DA}
\end{figure*}

For our final results, we first fit the ammonia abundance between -45 and 45 degrees independently at 1 deg steps, and then average the result over the chosen latitude range.  We adopt the value at 100 bar as the deep NH$_3$ abundance. We compile previous results for the averaged deep ammonia abundance as a function of pressure in \Cref{fig:DA} where the red line corresponds to the thermo-chemical equilibrium model (Model 1) and blue line to the stochastic model (Model 2), while the shading indicates the sensitivity of the retrieved results to the uncertainties in the retrieval. The 3-$\sigma$ profile for Model 1 converges at  $340.5^{+34.8}_{-21.2}$ ppm, where the uncertainties correspond to the 1-$\sigma$ and 10-$\sigma$ profiles. The final value corresponds to an enhancement over solar of $2.30^{+0.24}_{-0.14}$. We can see that our results confirm the results from the preliminary inversion of the temperature only measurement of PJ 1 \citep{Li2017} {($362^{+33}_{-33}$, $2.45^{+0.22}_{-0.22}$)} and are less than Galileo results \citep{Wong2004} and the VLA results \citep{dePater2019}, which are based on the Galileo results and are very insensitive to ammonia at these depths. 

\subsection{Orbit mean} 
\begin{figure*}[h]
\centering
\includegraphics[width=0.95\textwidth]{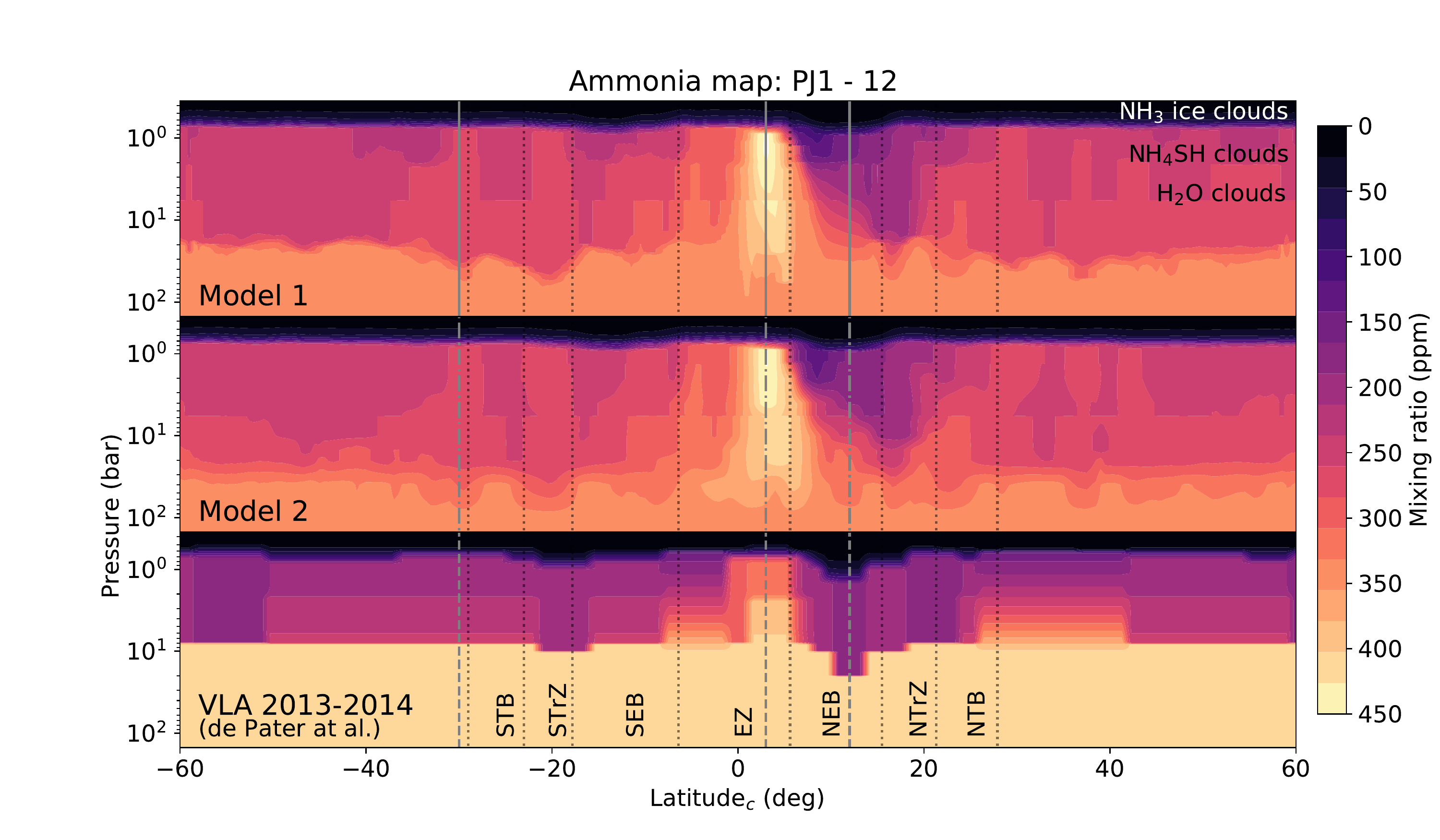}
\caption{Ammonia abundance map based on Juno's orbit average brightness temperature and limb-darkening with a 3-$\sigma$ fit uncertainty level. The map shows familiar features such as the column of enriched ammonia in the EZ and the depletion of ammonia in NEB, with smaller scale features at the mid-latitudes. Plotted on the same color scale is the original VLA map based on observations between 2013 and 2014 \citep{dePater2016,dePater2019}. The overall structure is very similar between the two maps, with the Juno data allowing to resolve more structure at depth and {the transitions in the atmosphere} due to its low frequency observations. The dotted lines indicate the latitude at which the zonal wind speed peaks, and with that denotes the difference between the zones and belts. The vertical gray lines correspond to locations of the atmospheric cross sections in \Cref{fig:NH3cut-PJmean}.} 
\label{fig:NH3-PJmean}
\end{figure*}

\begin{figure*}[h]
\centering
\includegraphics[width=0.95\textwidth]{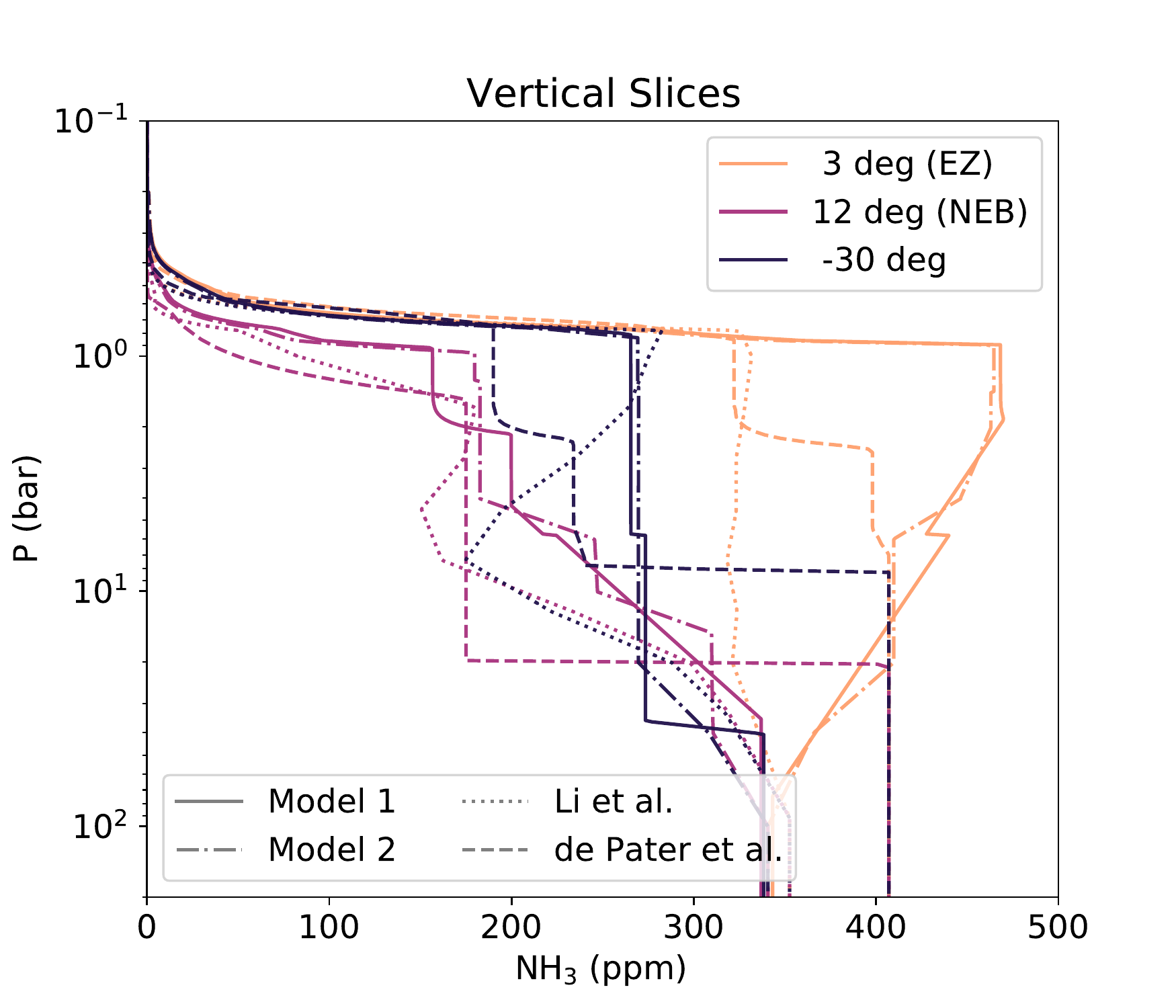}
\caption{{Comparison of the atmospheric structure at the locations indicated in \Cref{fig:NH3-PJmean}. The two models developed for this work agree well on the overall structure of the atmosphere despite different modeling frameworks.  The biggest difference in the ammonia distribution between our results and VLA results (dashed lines) \citep{dePater2016,dePater2019} is that we add more information about the fine-scale vertical stucture. Compared to the PJ1 results \citep{Li2017}, we find no strong evidence for the inversion at midlatitudes.} } 
\label{fig:NH3cut-PJmean}
\end{figure*}

With the deep ammonia constrained to a constant global value, we can turn our attention to obtaining a profile averaged over Juno's primary mission by fitting parameters to the average of the first two years of the mission between -60$^{\circ}$ and 60$^{\circ}$ (\Cref{tab:JunoCampagain}). Poleward of 60 degrees the spacecraft's larger distance from the planet reduces the resolution and increases contamination from the sidelobes, reducing the reliability of the data inversion.  The nine orbits are widely separated in longitude reflecting a global latitudinal structure, but we caution that the observations were taken over the course of a year, during which time the atmosphere underwent some large scale changes, such as the storm outbreak between the North Temperate Belt (NTB) and the North Tropical Zone (NTrZ) \citep{Sanchez-Lavega2017,dePater2019_2}, and the outbreak in the South Equatorial Belt (SEB) \citep[e.g., ][]{dePater2019_2}. In \Cref{fig:NH3-PJmean} we compare the average atmosphere based on the 3-$\sigma$ option to the results obtained from the VLA, which were taken between Dec.  2013 and Dec. 2014 \citep{dePater2016, dePater2019}. The VLA results are based on the brightness temperature at 21 different frequencies between 3 and 25 GHz, i.e., at a frequency resolution of 1 GHz; these data are sensitive to pressures $\lesssim$ 8-10 bar (and down to $\sim$20 bar in the NEB). In contrast, the Juno observations cover a much wider frequency range, from 0.6 to 22 GHz, at 6 distinct frequencies (Table 2), probing down to 100's of bars (see \Cref{fig:WF}). Although these data have a lower frequency resolution, there is further information in the Juno data by adding a second quantity to fit models: the variation of the limb-darkening coefficient with latitude. \\ 
The Juno observations {agree with} the overall structure of the planet that the VLA has seen at $P \lesssim$ 8 bar{, that is a depleted atmosphere above a well mixed region, while extending this analysis to greater depth.} The overall structure is characterized by alternating columns of enriched and depleted ammonia gas. The equatorial zone is highly enriched in ammonia, corresponding to the cold radio-temperatures around the equator. The Juno observations add fine scale structure, such as the increase of ammonia with altitude in the EZ at P $\lesssim$ 50 bar, the only location on the planet with such peculiar structure {(see also the yellow lines in \Cref{fig:NH3cut-PJmean})}. We can also observe the expansion of the equatorial column into the NEB at pressures between 2 and 20 bars. {The ammonia abundance in the EZ peaks high up in the troposphere and is enhanced over the deep ammonia abundance.} This feature ('ammonia enhancement factor') was instrumental to derive the water abundance in the EZ based on Channel 2 - 6 observations by \cite{Li2020}. The largest depletion of ammonia on the planet is in the adjacent NEB, where we fit a depletion of trace gases down to several tens of bars, quite similar to that derived from the VLA data, though slightly shifted northwards in latitude at larger pressures. At pressures of 40 bars, the depletion tapers off to the deep ammonia abundance. At higher latitudes the variation in brightness temperature with latitude becomes smaller resulting in a more homogeneous atmosphere, with most changes seen in the upper few bars,  similar to the VLA results \citep[e.g., ][]{dePater2016}.

\begin{figure*}[h]
\centering
\includegraphics[width=\textwidth]{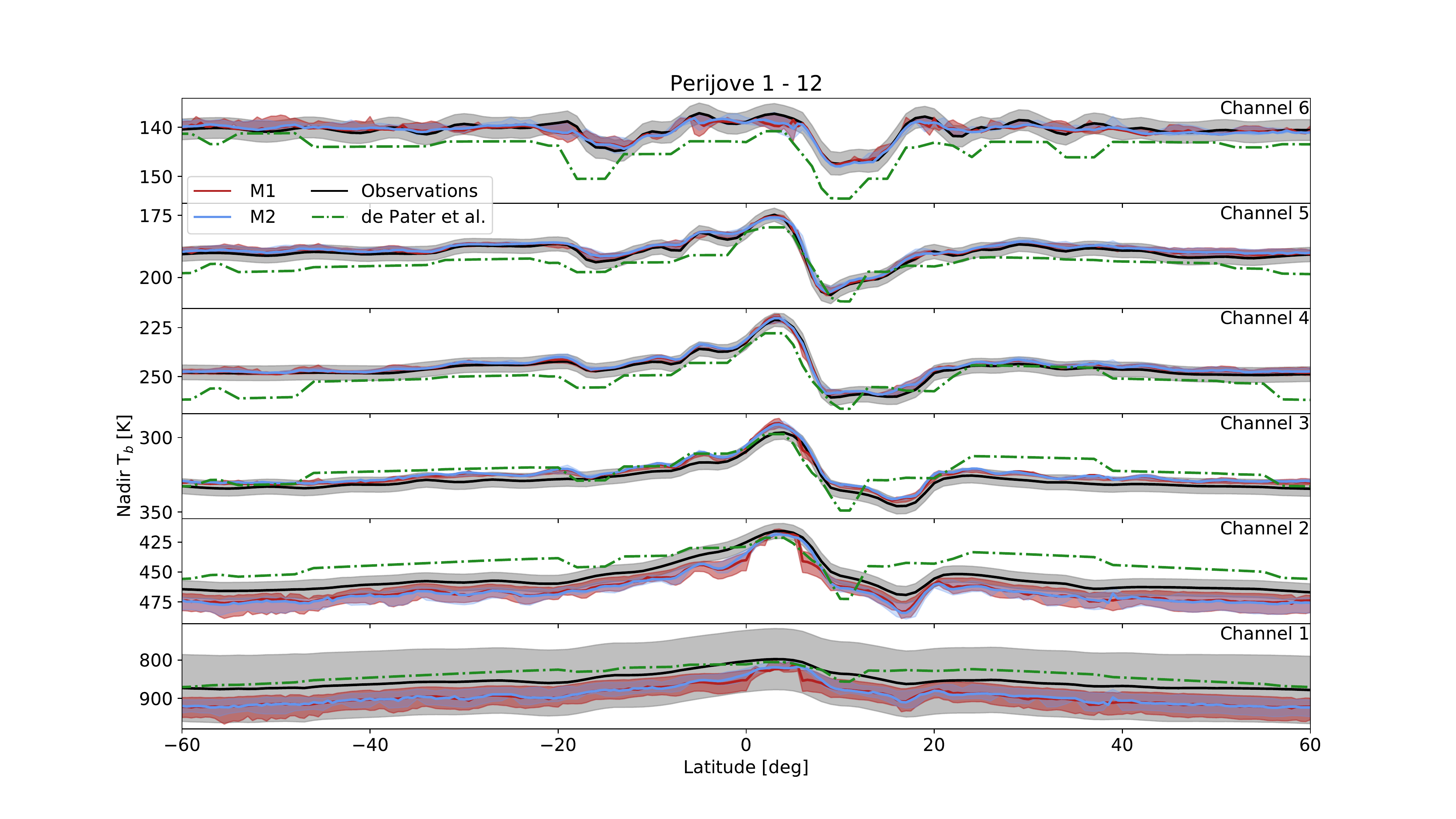}
\caption{Forward modeling results based on the best fit distribution of ammonia trace gases compared to the Juno observations, plotted against latitude. The Juno nadir brightness temperatures are the solid black lines and with their uncertainties represented by the shading.  The red and blue lines correspond to temperatures based on the ammmonia distributions as shown in \Cref{fig:NH3-PJmean}. The shading of the red and blue lines indicates the sensitivity of the results to the uncertainties (n=1,10). The green line is the forward model based on the VLA data \citep{dePater2019}. Overall, our forward models fit the brightness temperature well within the uncertainties.}
\label{fig:TBfit_PJmean}
\end{figure*}

\begin{figure*}[h]
\centering
\includegraphics[width=\textwidth]{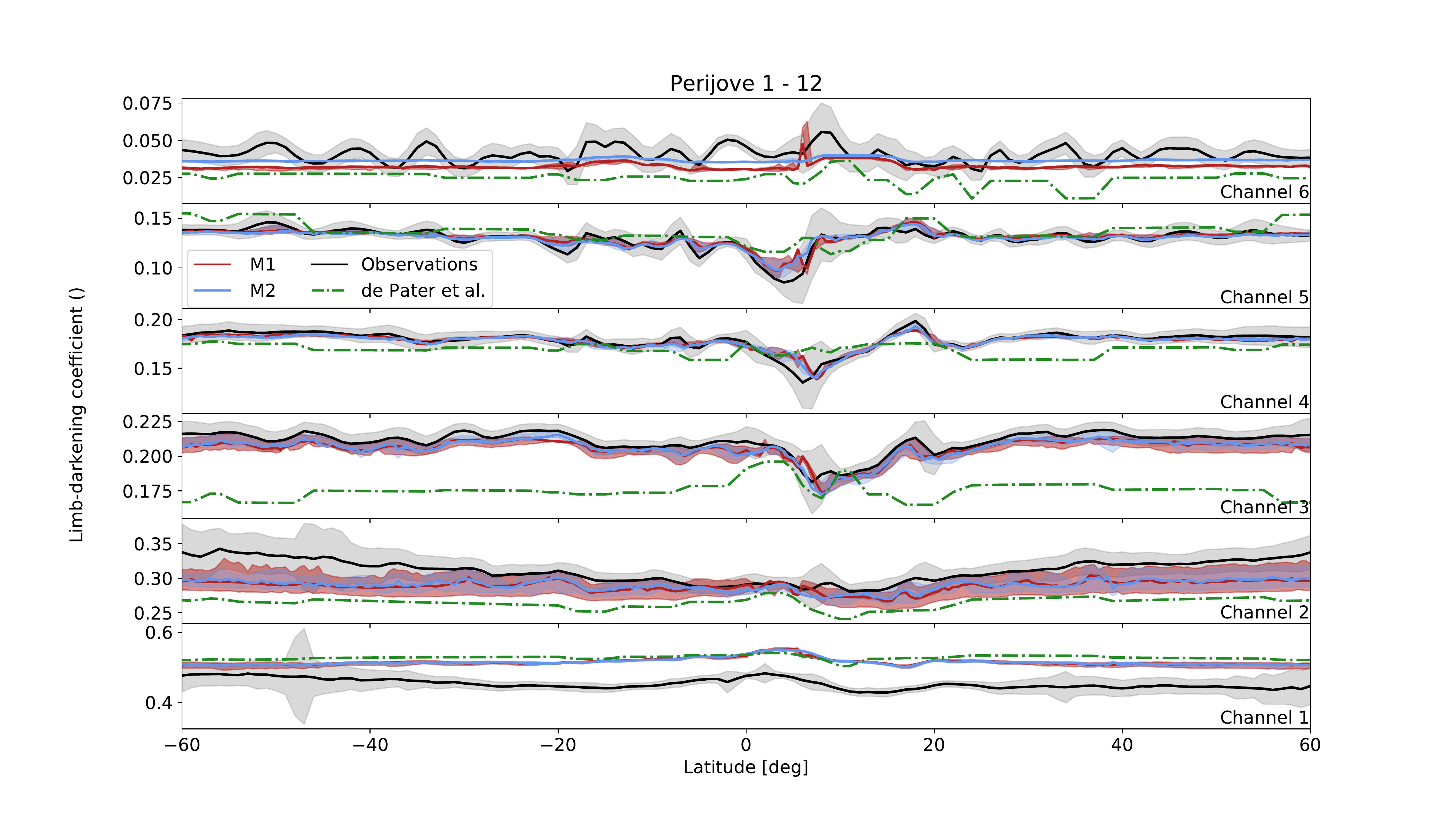}
\caption{Same as \Cref{fig:TBfit_PJmean} but for the limb-darkening coefficient. While we can fit the overall structure of the atmosphere within the uncertainties, the solutions for Channel 2 all result in less limb-darkening compared to the observations. The solution closest to the observations (upper end of the shaded area) corresponds to the retrieval that creates a larger mismatch for the brightness temperature. The results that best fit the brightness temperature (10-$\sigma$) have the largest misfit for the Channel 2 limb-darkening, and the 1-$\sigma$ results fit the limb-darkening best.}  
\label{fig:pfit_PJmean}
\end{figure*}

We assess the goodness of our fit by computing the brightness temperature and limb-darkening coefficient for the given ammonia distribution in \Cref{fig:NH3-PJmean}. Next to the best-fit results that is the basis for our interpretation (i.e., the 3-$\sigma$ option), we also show the sensitivity of the our retrieved atmosphere for the other combinations of uncertainty (n=1 and n=10) in Figures \ref{fig:TBfit_PJmean} and \ref{fig:pfit_PJmean} { through the use of shading. The width of the shading for Channel3 - Channel 6 is very small which indicates that the uncertainty matters little for the higher frequencies in fitting the observations, and our best-fit distribution can explain the observed brightness distribution within the Juno MWR uncertainties.} The biggest impact of our confidence in the uncertainties appears in Channel 2, where the various combinations trend to a more depleted deep atmosphere (n=1, 1-$\sigma$) or a more enriched deep atmosphere (n=10, 10-$\sigma$), reflected by the increased width of the shading. The brightness temperature fit in Channel 2 is anti-correlated with the limb-darkening fit in Channel 2. The temperature fits best when using a large scaling factor $n$ (such as 10) for the fit uncertainty, in which case the brightness temperature drives the retrieval. A scaling factor of n=1, on the other hand, means we fit the limb-darkening at the cost of brightness temperature. We also show the brightness temperature (green line) for the distribution determined by the VLA \citep{dePater2019}. We can see that the VLA observations, while not being in perfect agreement reflect the overall structure of the atmosphere pretty well. The VLA model (based on the lower end of the Galileo Probe data \citep{Wong2004})  is generally too cold for Channel 1 to Channel 3, due to its lack of sensitivity to radiation from the deeper layers in the planet (see \Cref{fig:WF}). \\ 

While latitudinal structure in the limb-darkening coefficient is more chaotic, we still manage to fit the overall structure of the planet to first degree. The large features around the equator and the neighboring belt are mostly captured by the fit, only the smaller diversion at higher latitudes tend to not be captured well by our fit. This figure shows the additional value of the limb-darkening coefficient in constraining the atmosphere. While the brightness temperature for the VLA distribution was comparable to the Juno observations (\Cref{fig:TBfit_PJmean}), the limb-darkening coefficient for Channels 2 and 3 does not match, and indicates that the atmosphere is well-mixed at altitudes much deeper than the 10 bar level that was inferred from the VLA data. Higher up in the atmosphere the VLA observations agree well with the Juno observations; near the ammonia ice cloud, the VLA shows larger latitudinal variations than the Juno data, which likely results from the higher frequency resolution. While we do not fit Channel 1, it is still interesting to see that all models produce similar results, with much larger coefficients than observed. \\

\subsection{Perijove 1} 
\begin{figure*}[h]
\centering
\includegraphics[width=\textwidth]{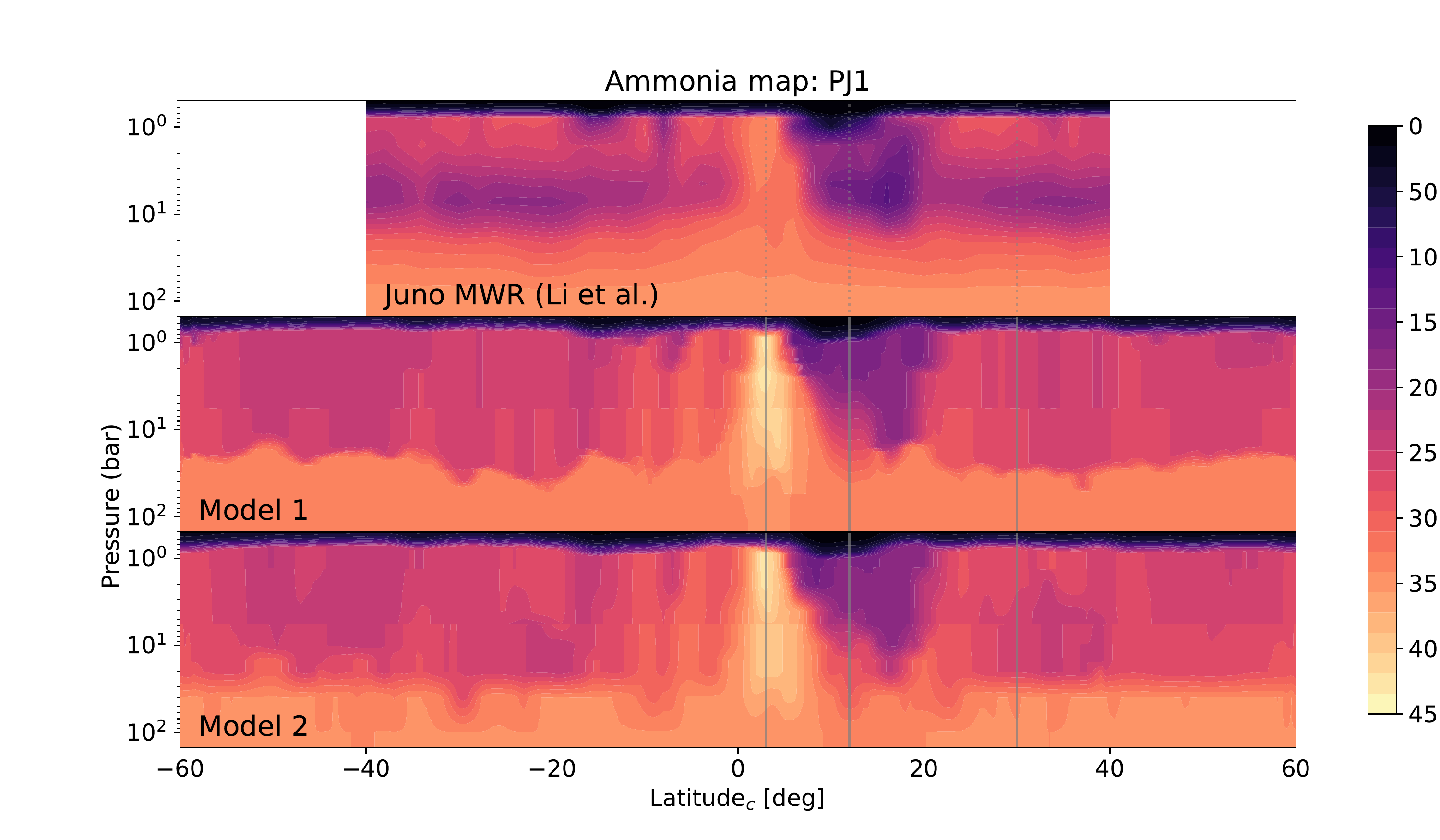}
\caption{Comparison of ammonia abundance map for PJ1, based on the preliminary retrieval \citep{Li2017} and the best-fit model in this paper. The overall structure looks very similar, with the biggest difference that we do not find evidence for the ammonia inversion centered around 6 bars in the earlier retrieval as obtained by \citet{Li2017}. } 
\label{fig:PJ1NH3map}
\end{figure*}

We now turn our focus to compare our model to results published by \citet[between 40$^{\circ}$S and 40$^{\circ}$N]{Li2017}. \Cref{fig:PJ1NH3map} shows our map (3-$\sigma^{fit}$) for PJ 1 together with the results from \citet{Li2017}; the latter were based solely on the brightness temperature \cite{Li2017}. We can reproduce the enhancement of ammonia at the equator and the depletion in the North Equatorial Belt. The \citet{Li2017} map shows that latitudinal variations are constrained to the top few bars of the atmosphere, above an ammonia inversion, i.e., minimum in the ammonia abundance at all latitudes, with a well-mixed deeper layer. In contrast, our model converges to a solution that monotonically increases the ammonia abundance with increasing pressure. The depth that we can trace this variation to, however, is deeper than the \citet{Li2017} results. 

\begin{figure*}[h]
\centering
\includegraphics[width=\textwidth]{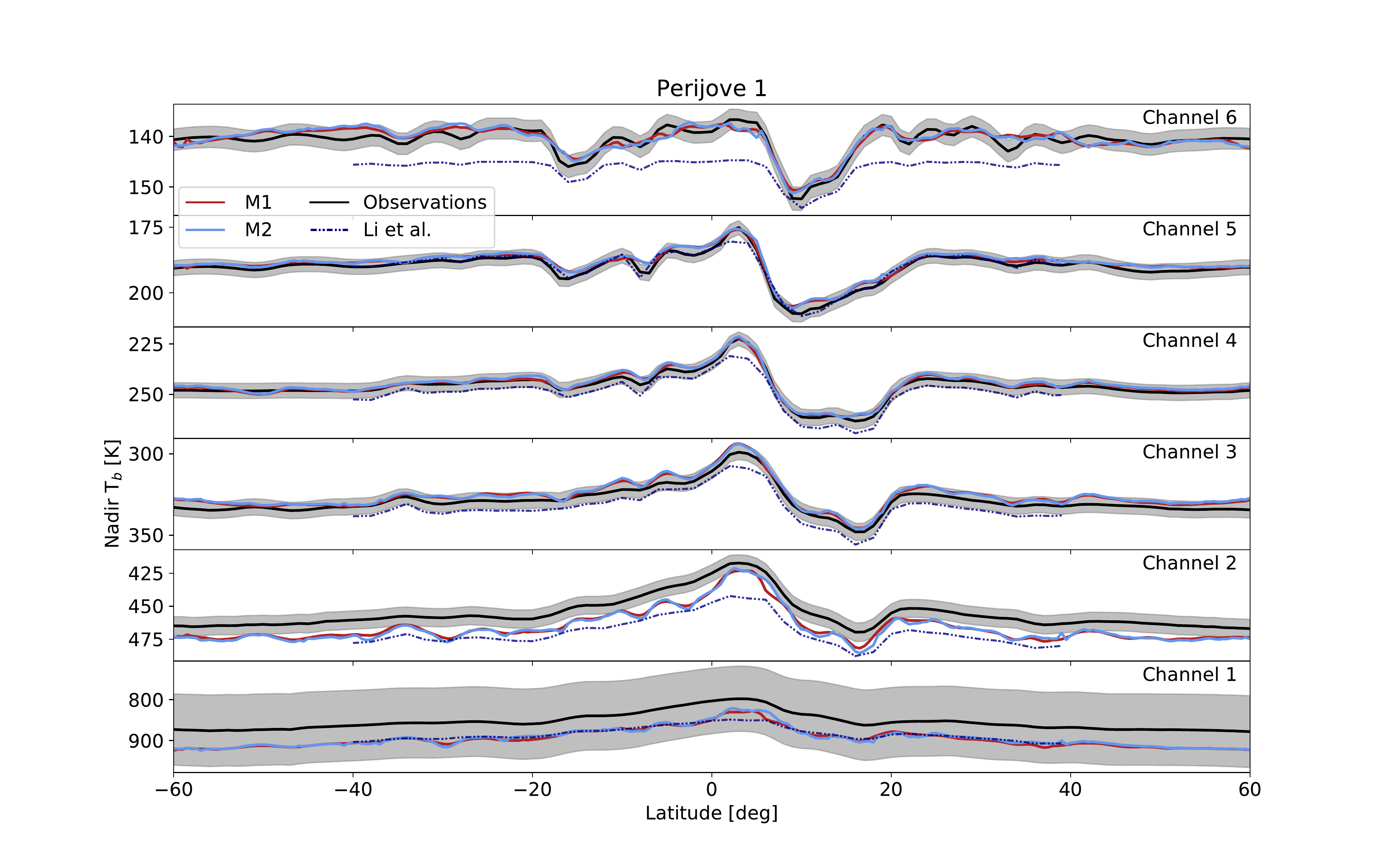}
\caption{Same as \Cref{fig:TBfit_PJmean} but for the PJ1 observations. The MWR brightness temperature observations are given by the solid lines, their uncertainty is given by the shaded region. The red and blue solid lines are the brightness temperatures for the abundances as plotted in \Cref{fig:PJ1NH3map}. The dash-dotted line corresponds to the forward modeling results based on the ammonia distribution as published by \citet{Li2017}.  We note that while model the  structure of the atmosphere differently, both models are close to the uncertainty.}
\label{fig:TBfit_PJ1}
\end{figure*}

\begin{figure*}[h]
\centering
\includegraphics[width=\textwidth]{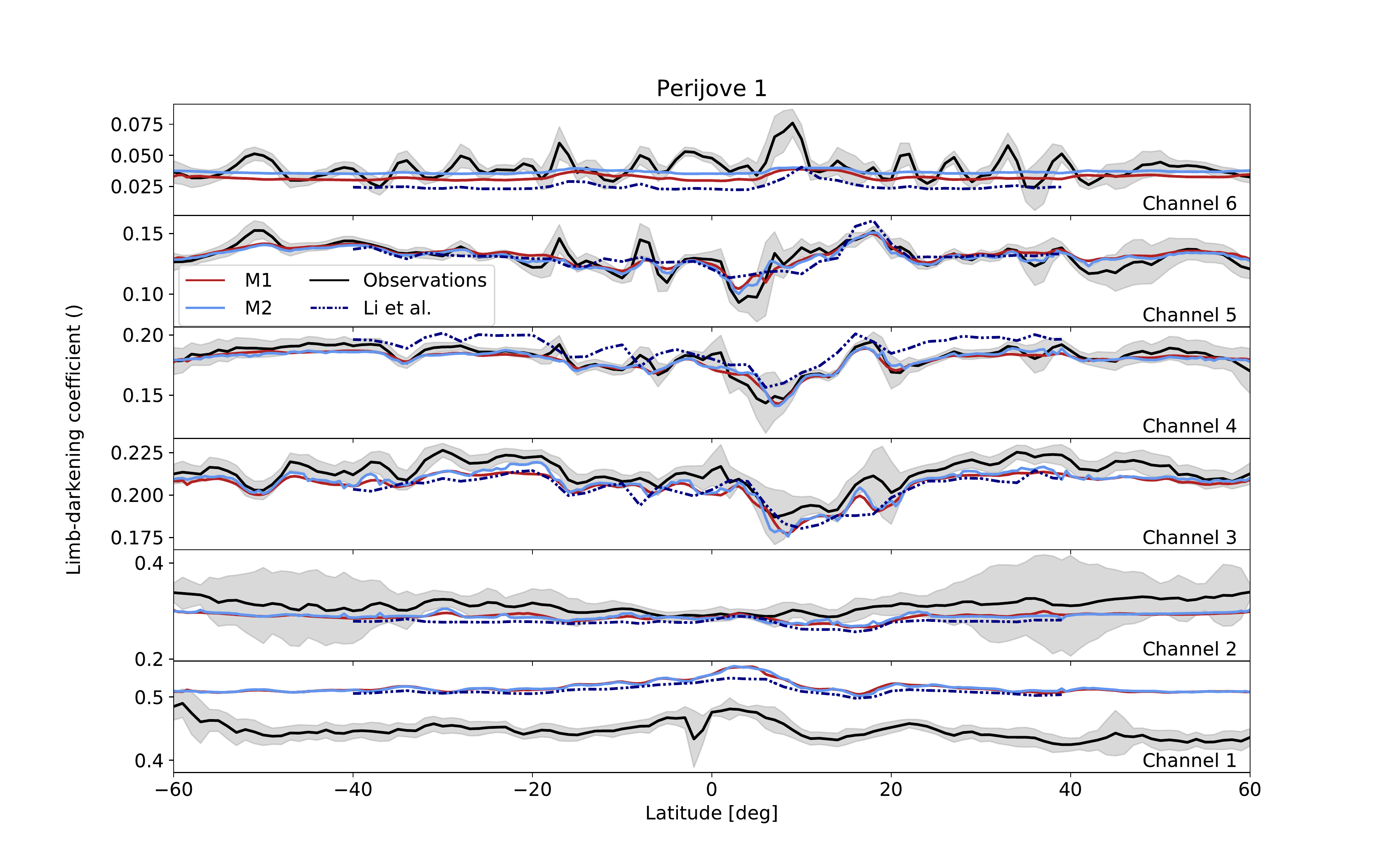}
\caption{Same as \Cref{fig:TBfit_PJ1} but for the limb-darkening coefficient.  }
\label{fig:pfit_PJ1}
\end{figure*}

In Figures \ref{fig:TBfit_PJ1} and \ref{fig:pfit_PJ1} we forward model the ammonia distribution in PJ1 using our radiative transfer code and compare it to the observations for PJ1 (our models: blue and red solid lines; \citet{Li2017}: blue dash-dotted line). The four highest frequencies corresponding to the upper layers of the atmosphere are well fit within the uncertainties, and our results in all channels match the data better than the results in \citet{Li2017}. This highlights how degenerate inversions based on radio observations can be, where quite different ammonia profiles can approximate the observations well.

\subsection{Combined VLA 2013-2014 and Juno PJ1-12 results}
\begin{figure*}[h]
\centering
\includegraphics[width=\textwidth]{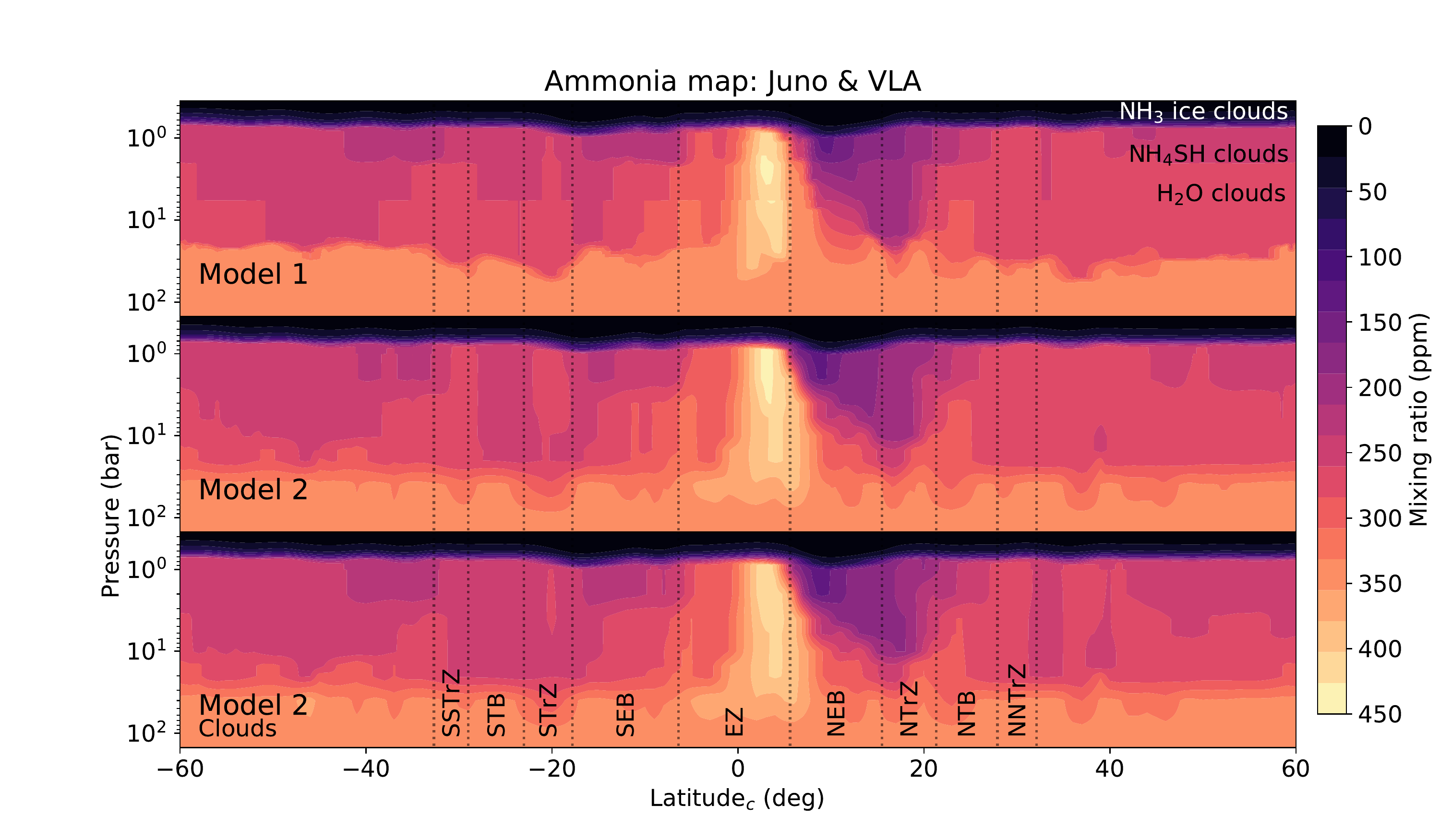}
\caption{{Ammonia distribution based on the combined Juno and VLA observations. The top panel shows the distribution based on the thermo-chemical equilbrium model, the middle panel is based on the stochastic model, while the lower panel shows the effect of clouds onto the radio spectrum.}}
\label{fig:NH3-Comb_v1}
\end{figure*}

With the updated deep ammonia abundance based on the Juno data, we can now {add further observational constraints using the longitudinally averaged} profiles of the VLA data taken $\sim$2 years prior to the Juno orbit insertion \citep{dePater2016,dePater2019}. \Cref{fig:NH3-Comb_v1} shows our ammonia abundance results, {as retrieved from the combined VLA/Juno frequency coverage. Due to the different viewing directions of Juno (nadir) and the VLA (looking from Earth) we chose not to convert the VLA observation into nadir viewing\textbf{, which would require assumptions about the limb-darkening.} Instead, we model the atmosphere as seen from Earth, that is zonally averaged latitude slices between 3 and 24 GHz separated by 1GHz where the emission angle is determined by the observing geometry (see also \citep{dePater2019,dePater2019_2}). This adds further constraints on the atmosphere down to $\sim$ 10bar level, at deeper levels Juno Channel 1 to Channel 3 observations constrain the ammonia distribution. Both our models converge to a solution that shows an atmosphere that is depleted in trace gases down to $\sim$ 20-30 bar, above which the atmosphere shows variations between zone-belt variation on the order of 50 ppm. The bottom panel of \Cref{fig:NH3-Comb_v1} shows an atmosphere where we include the effects of scattering and absorption by cloud particles in the radiative transfer model based on parameterization by \citet{dePater1985,dePater2005}. There are many unknowns regarding the clouds including the cloud particle density, the imaginary index of refraction and their vertical distribution. Our cloud parameterization represents the end member case by assuming large values for the cloud density. The goal is to show that the overall structure of the atmosphere does not change dramatically, rather it modifies the small scale structure. Most notably, it removes ammonia higher up in the atmosphere since cloud particles provide extra absorption. The largest effect is visible in the equatorial zone, where ammonia does not increase substantially as compared to the upper two panels which ignore cloud opacity. We model the impact of clouds on the equatorial structure in some more detail in Section \ref{ssec:ezinv}.}

\section{Discussion} \label{sec:discussion} 
{In this section we first discuss the uncertainties in the retrieved atmosphere by showing results of MCMC fits to only the Juno MWR observations, and then we focus on the dynamics of the atmosphere based on the retrieved parameters and what inferences we can make from the radio data. Two particularly interesting features emerge in Jupiter's atmosphere: the inversion in the equatorial zone, and the structure at mid-latitudes where \citet{Li2017} found an inversion at $\sim$6 bar. These features are studied in more detail in Section \ref{ssec:ezinv} and \ref{ssec:ML} by combining the Juno and VLA data. } 

\subsection{Uncertainties for dynamical tracers}\label{sec:uncertainties} 
\begin{figure*}[h]
\centering
\includegraphics[width=\textwidth]{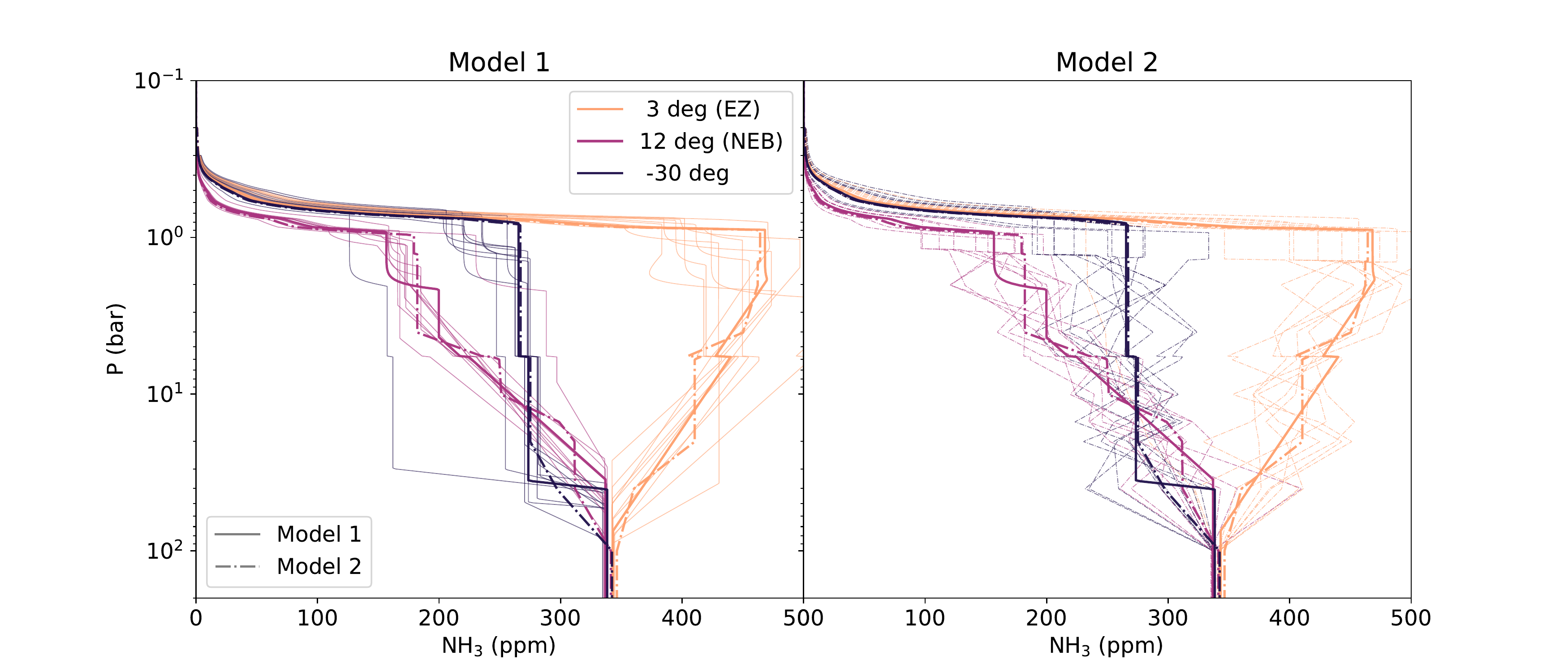}
\caption{Slices through the atmosphere at the same locations as indicated in \Cref{fig:NH3-PJmean}. We plot additionally here 10 draws from the MCMC results to show the spread in solutions that are consistent with the observations. The left panel shows the solutions for our perturbed thermo-chemical equilibrium model while the right hand side shows the results for our stochastic model. We can see that overall structure is reproduced well by the MCMC draws, indicating the confidence in the retrieved solutions.}
\label{fig:NH3cut-MCMC}
\end{figure*}

{In order to assess the uncertainties in our results, we performed a MCMC retrieval based on just the Juno data on selected slices of the atmosphere in the EZ and NEB. MCMC results add two more layers of information: uncertainties in and potential correlations between parameters. In \Cref{fig:NH3cut-MCMC} we overplot our results from the MCMC result onto the optimizer results as shown in \Cref{fig:NH3cut-PJmean}, where the spread between the various MCMC draws shows the confidence in the optimizer solution.  The left panel shows the variations for our therm-chemical equilibrium approach, while the right panel shows the results for the stochastic model. Despite variations on the order of 50 ppm between the MCMC draws, the overall structure is reproduced in the MCMC results indicating that the optimizer results are well-constrained by the observations and representative. Both models indicate an increase of ammonia abundance with altitude in the equatorial zone. } 

\begin{figure*}[!h]
\centering
\includegraphics[width=0.75\textwidth]{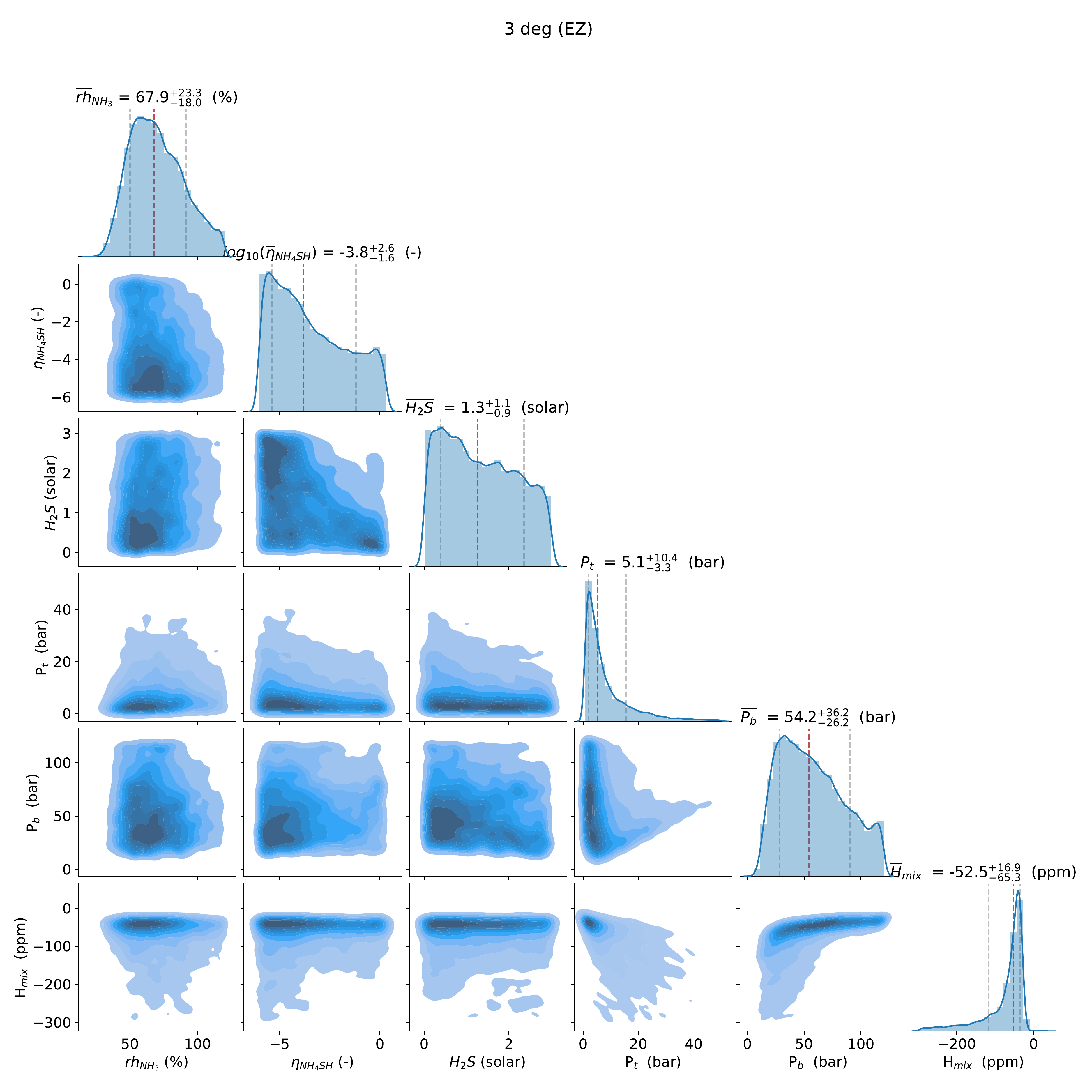}
\caption{Uncertainties for and correlations between the retrieved parameters in the Equatorial Zone. }
\label{fig:corner_EZ} 
\end{figure*}

\begin{figure*}[!h]
\centering
\includegraphics[width=0.75\textwidth]{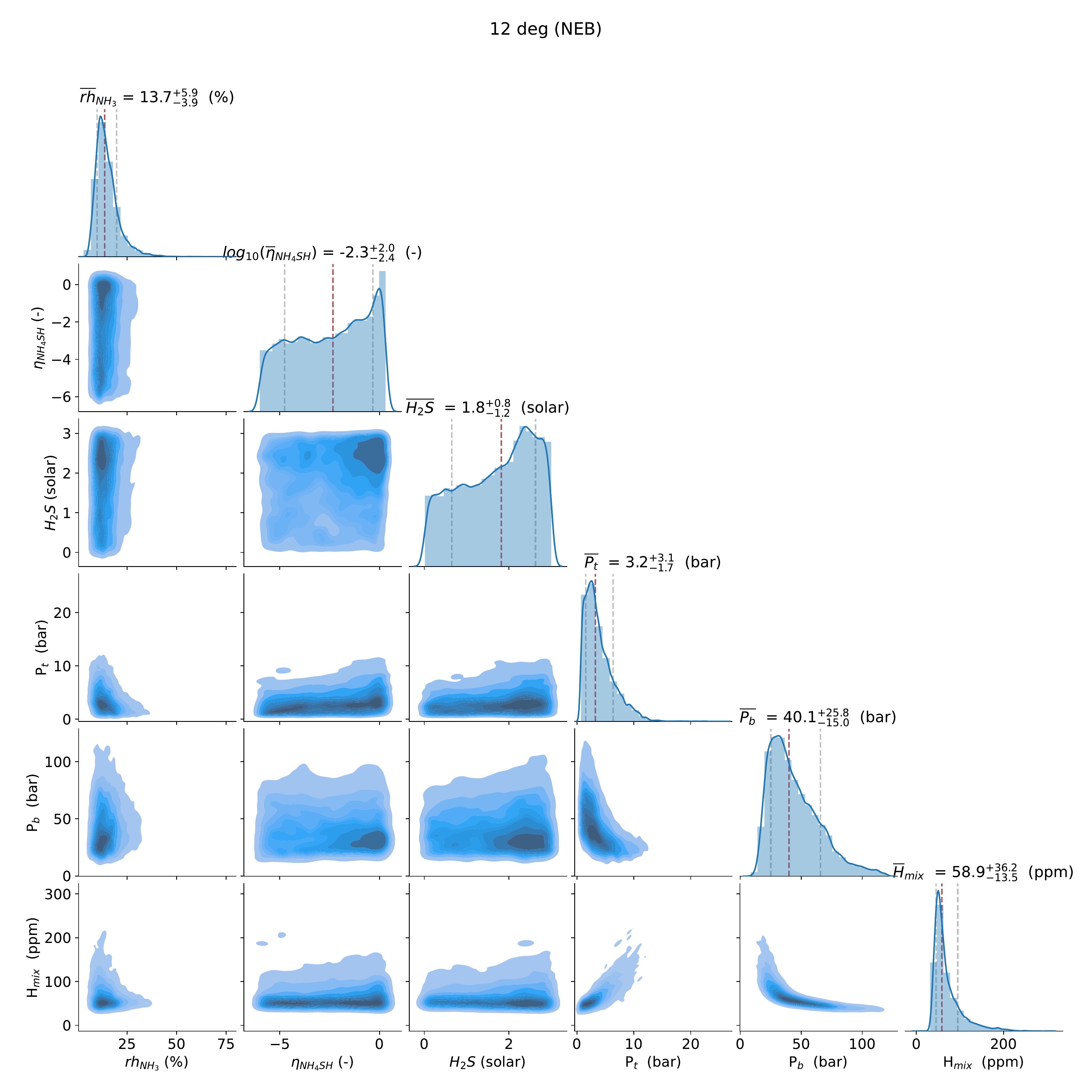}
\caption{Same as Figure \ref{fig:corner_EZ} but for the North Equatorial Belt. }
\label{fig:corner_NEB}
\end{figure*}

 The diagonal elements of the corner plots in  \Cref{fig:corner_EZ,fig:corner_NEB} show the distribution of solutions for the given parameters with the error bars corresponding to the mean, and the 0.16 and 0.84 quantiles. The off-diagonal elements indicate the correlation between parameters. Starting at the top of the atmosphere, we can see that the relative humidity in the ammonia ice cloud is well constrained. We find a high relative humidity in the EZ, and low humidity in the NEB, which is in line with the thick white cloud deck seen at visible wavelengths around the equator and the lack of the white ammonia clouds in the NEB. Such differences in relative humidity were first suggested by \citet{Klein1978}, and later verified from VLA observations \citep[e.g., ][]{dePater1986,dePater2016,dePater2019}.

Sitting below the ammonia clouds is the postulated $NH_4SH$ cloud, that forms around 2-3 bar. From our retrievals we find that both the pressure at which the clouds forms and the amount of $H2S$ in the atmosphere is poorly constrained. This is mostly due to the fact that we have limited coverage at the pressures where the clouds supposedly form due to the wide spacing of the MWR frequencies.\\ 

We define the region where mixing occurs by three parameters (see \Cref{eq:dnh3dp}): the gradient ($H_{mix}$)  between a bottom pressure ($P_b$) and a top pressure ($P_t$). The standing column of ammonia around the EZ leads to poor constraints on the mixing region, however all solutions in the MCMC run support an increase of ammonia with altitude (see more in Section \ref{ssec:ezinv}). In contrast, the mixing range  in the NEB occurs deeper in the planet, and both $P_t$ and $H_{mix}$ are much better constrained.

\subsection{Atmospheric Dynamics}
\begin{figure*}[h]
\centering
\includegraphics[width=\textwidth]{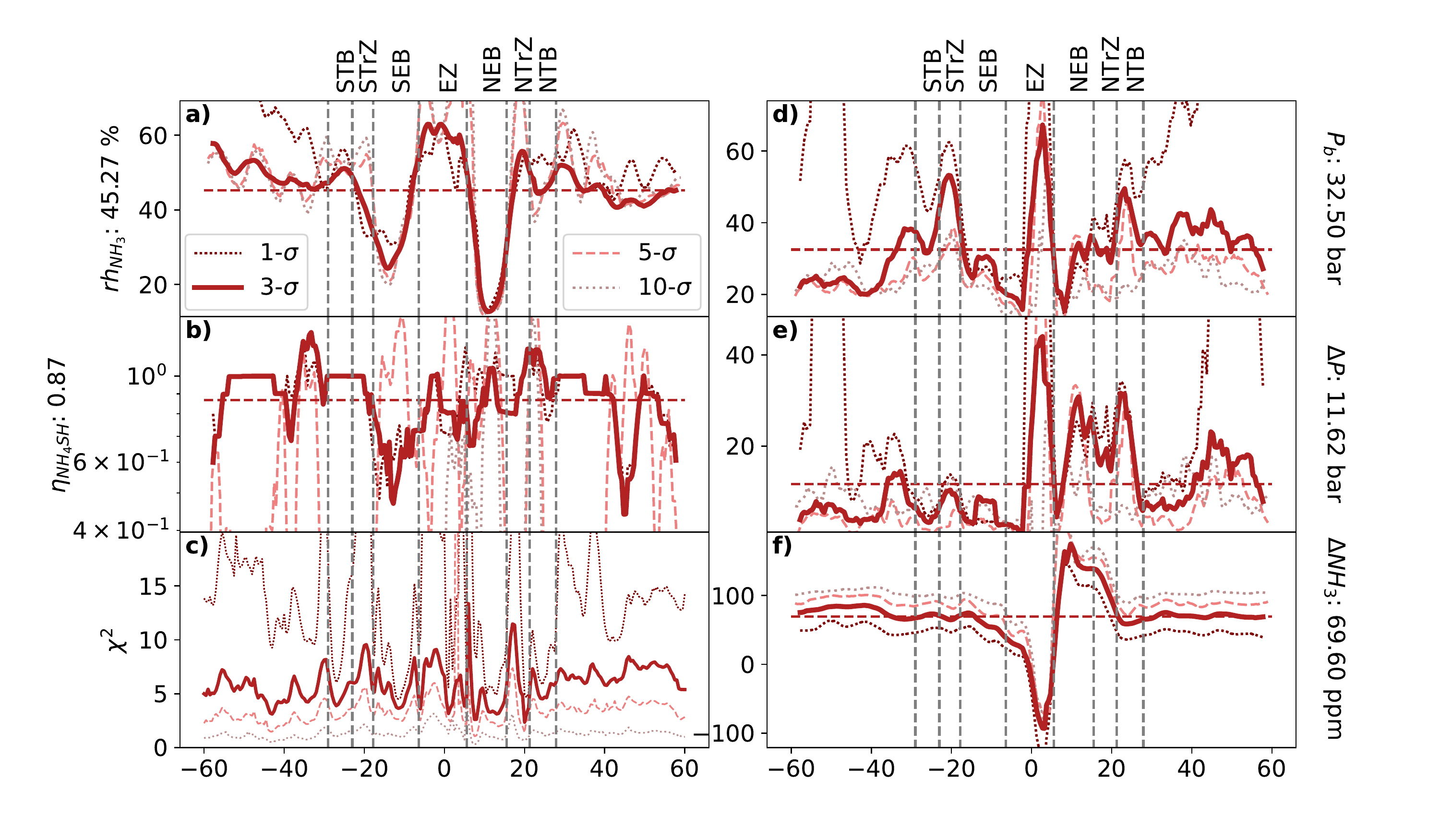}
\caption{Retrieved atmospheric parameters as a function of latitude (with the individual zones and belts denoted by the vertical dashed lines) and for various combinations of uncertainty (different colors and linestyles). \textbf{Left}: a) Relative humidity of ammonia gas, b) Cloud formation efficiency, c) The reduced - ${\chi}^2$ value for the retrieved atmosphere. \textbf{Right}: d) Bottom pressure of mixing region, e) Pressure range over which mixing occurs, f) Absolute change in ammonia abundance in the  mixing region.     }
\label{fig:PJmean_params}
\end{figure*}

In the following we use our retrieved parameters for the ammonia distributions as a proxy for atmospheric dynamics. Our approach to combine measurements into 1 deg bins and retrieve the parameters per bin results in a large spread between adjacent latitudes, while the spatial resolution of the individual observations is coarser than 1 deg (\Cref{fig:PJ3_map}). To improve the signal we average the parameters over 5 degrees of a latitude-rolling-average to show trends across the zones and belts. \Cref{fig:PJmean_params} shows the retrieved best-fit parameters for the set of uncertainty combinations. 

\subsubsection{Relative humidity} 
We obtain an increased relative humidity in the equatorial zone along with a decreased relative humidity in the adjacent NEB and SEB compared to other latitudes. Moving further towards the poles, the signal of the South Tropical Zone and North Tropical Zone can be seen by the increase in relative humidity until the relative humidity settles to a stable mean below 50\% (see panel a) in \Cref{fig:PJmean_params}). The Juno beam is large in comparison to cloud features, so that we interpret a value of less than 100\% saturation as a mixture of clouds and cloud-free regions.  

\subsubsection{NH$_4$SH cloud formation}
The MCMC runs revealed that the actual pressure where $NH_4SH$ cloud formation starts is poorly constrained, however, our 3-$\sigma$ models return a consistent value close to unity for $\eta_{NH_4SH}$. A value close to 1 indicates that the onset of cloud formation happens as predicted by the thermo-chemical equilibrium models, while a value less than one indicates that the reaction is happening at lower pressures than predicted. The nominal pressure at which $NH_4SH$ is the more favorable state of the trace gases is nominally around 2.3 bars for a value of 1, whereas a value of 1e-3 pushes that cloud formation to a pressure of 1.4 bar. This can either mean that the process is a lot less efficient than postulated \citep{Weidenschilling1973}, for example if there are updrafts that "supercool" the gas before NH$_4$SH formation takes place. The disagreement between the various linestyles in panel b) of \Cref{fig:PJmean_params} however indicates that this result depends much on the choice of uncertainties. Alternatively, this could also reflect a lower abundance of H$_2$S, which we only constrain indirectly. { The retrievals for the EZ and the NEB (see \Cref{fig:corner_EZ,fig:corner_NEB}) show that H$_2$S abundance (EZ = 1.3$^{+1.1}_{0.9}$ solar, NEB = 1.8$^{+0.8}_{-1.2}$ solar) in both regions is lower than the ammonia abundance (2.30), however it is poorly constrained as seen by the large spread.}

\subsubsection{Mixing regime} 
Juno's MWR covers a wide frequency range at the expense of spectral coverage, resulting in very few constraints on the deeper atmosphere. Our thermo-chemical equilibrium model prescribes a single gradient ($H_{mix} \equiv \frac{\delta NH_3}{\delta ln(P)}$) starting from a bottom pressure ($P_b$), over a certain pressure range ($\Delta P$). For clarity we report on the absolute change in the ammonia abundance ($\Delta NH_3$) over the mixing regime, as the gradient itself is a lot less constrained, {but the product of the mixing over a certain pressure region is well constrained.} The mean bottom pressure at which we constrain the mixing to start is around {30} bars, but with significant difference between the tropical region and the mid latitudes (see panel d in \Cref{fig:PJmean_params}). {The pressure range over which mixing occurs is consistent in the extra-tropical region where the mixing occurs over a rather small pressure range ($\lesssim$ 5 bars), as shown in panel e of \Cref{fig:PJmean_params}. This mixing is very consistent across uncertainties, and much more localized in pressure than the results shown in \citet{Ingersoll2021}, which derived a map based on both limb-darkening and nadir brightness temperature information. }\\ 

{When we take a closer look at the structure where the atmosphere transitions from depleted to enriched ($\sim$ 20 bars) in \Cref{fig:NH3-Comb_v1}, a pattern emerges. The bottom pressure at which the atmosphere approaches the level of the global deep ammonia abundance, shows a persistent pattern between zones and belts. The zones while enriched in ammonia higher up in the atmosphere, are depleted deeper in the atmosphere. This pattern was also picked up by \citet{Fletcher2021}, who referred to this as the jovicline, akin to the thermocline in terrestrial oceans. The presence of this pattern supports the existence of a vertically-stacked circulation cell \citep{Showman2005,Duer2021}. The difference in at what pressure the atmospheres are well mixed between zones and belts is on the order of 10 bar as shown in panel d of \Cref{fig:PJmean_params}. }

Interestingly, the mixing occurs at levels well below the water clouds ($\sim$6 bar), which was long thought to be the main interface between vertically-stacked circulation cells, inhibiting convection from below \citep{Showman2005,dePater2019_2}. 
Our results confirm that the atmosphere is not well mixed below the cloud levels (NH$_3$, NH$_4$SH, solution cloud), as inferred as early as the 1980s when radio measurements showed the presence of a radio-hot NEB, which was interpreted as a depletion in the ammonia gas abundance \citep{dePater1986}. VLA measurements showed this depletion to extend down to the $\sim$20 bar level \citep{dePater2019}, and we show that it extends even deeper.\\

Despite its large size, and the inherent computational problems with modeling planets this size, \cite{Young2019,Young2019p2} have managed to simulate the dynamics of Jupiter, and found an ammonia depletion down to $\sim$8 bars, with the exception of the EZ, implying that dynamics may play a large role in shaping the atmosphere of Jupiter. Recently, microphysics has been invoked to explain the depletion of trace gases by trapping ammonia in water ice and raining it out \citep{Guillot2020}. It remains to be seen if this process can deplete an atmosphere down to tens of bars, i.e., well below the melting point of water ice. \\

While the mid-latitudes seem to have a very consistent signal for the mixing region, with $\Delta NH_3 \sim$ 100 ppm (panel f in \Cref{fig:PJmean_params}), the EZ and the NEB show the largest deviation from that number. Most notable, the EZ shows a reversal in the sign of the gradient, which results in an increase in the ammonia abundance with altitude, while the NEB shows the biggest gradient. 

\subsection{Equatorial Zone structure}\label{ssec:ezinv}
\begin{figure*}[!h]
\centering
\includegraphics[width=0.75\textwidth]{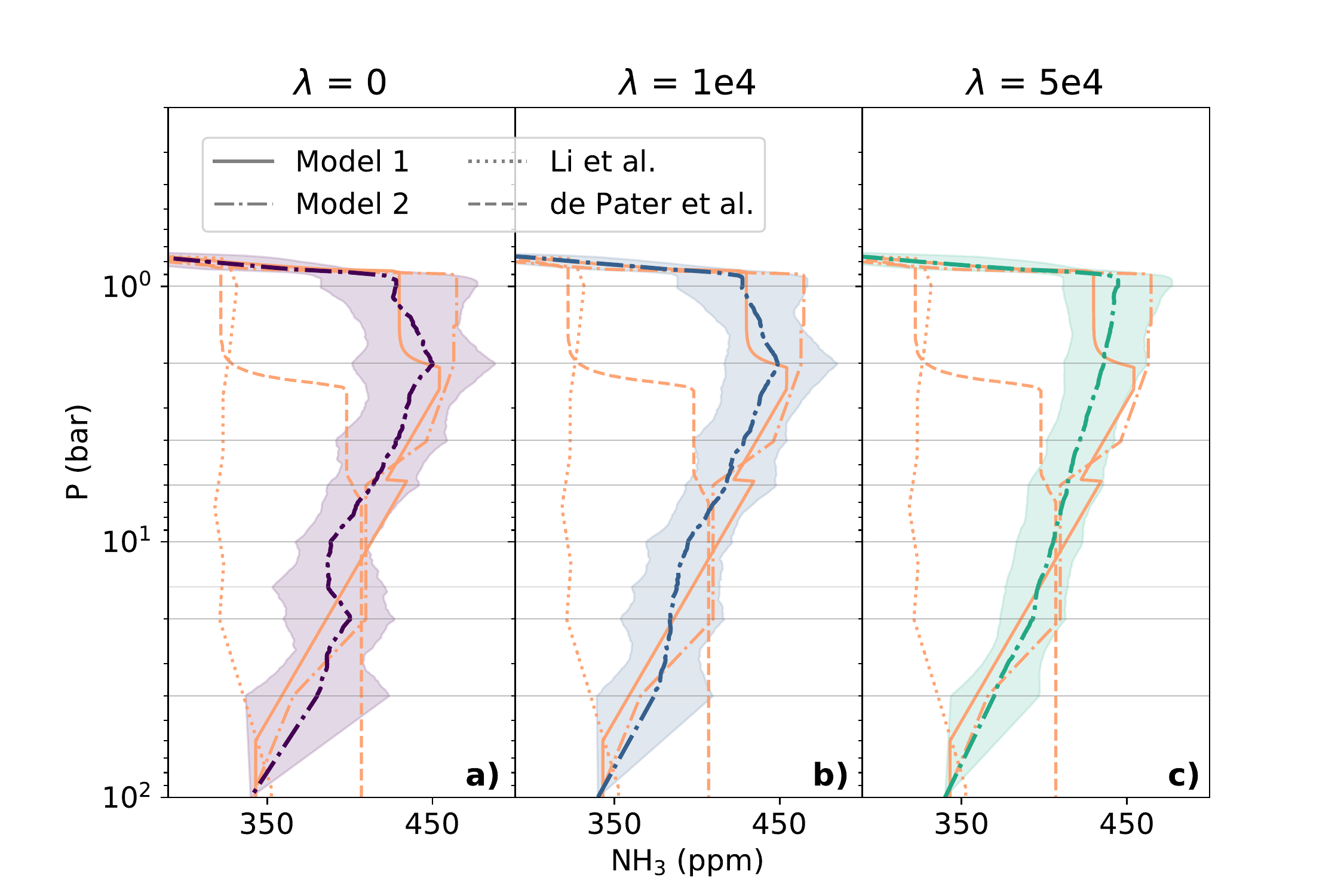}
\caption{Effect of regularization constrain on the ammonia abundance profile in the EZ when fitting to the combined VLA and Juno data. The yellow lines show the results of various different retrievals (and are the same as in \Cref{fig:NH3cut-PJmean}), while the horizontal lines are the pressure nodes between which the gradient can vary. The three different panels show the effect of the vertical regularization as described in Section \ref{ssec:Smodel}, where the shading corresponds to the 0.16 and 0.84 quantiles from the MCMC runs. The left panel has the regularization turned off ($\lambda=0$ in \Cref{eq:C2}), with a mild regularization constraint in the middle and a moderate constraint on the right. While the regularization constraint changes how smooth the profile is with altitude, it changes very little about the overall structure.}
\label{fig:EZ_regularization}
\end{figure*}

\begin{figure*}[!h]
\centering
\includegraphics[width=0.5\textwidth]{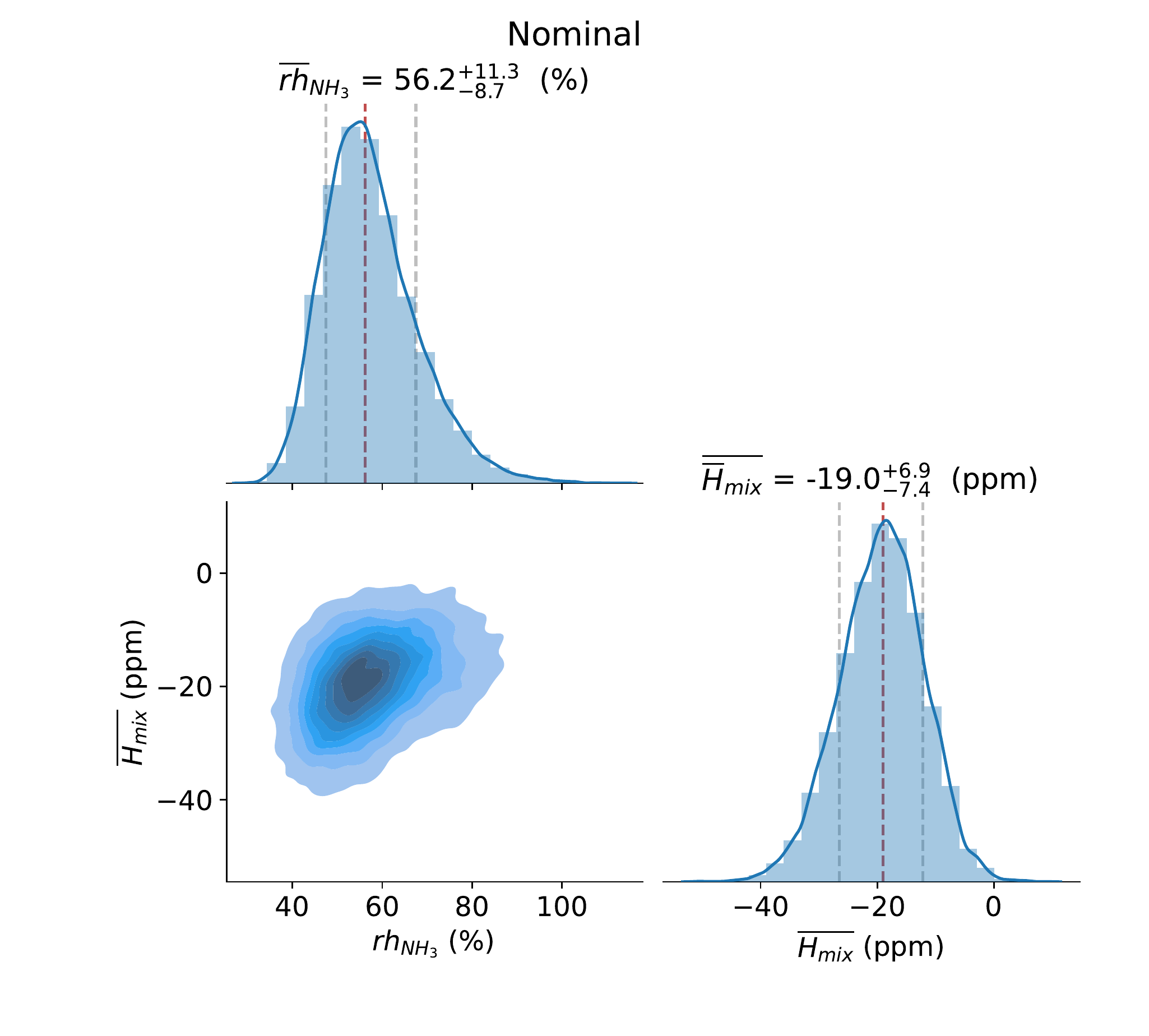}
\caption{{Corner plots for the nominal model (panel b in \Cref{fig:NH3cut-EZmodels} where we show the average between the 8 mixing gradients in the atmosphere between 100 and 1 bar. The combined VLA and Juno retrieval indicates a lower humidity in the EZ compared to the pure Juno retrieval (see \Cref{fig:corner_EZ}), but the mixing gradient clearly shows a well-defined increasion of ammonia abundance with altitude.}}
\label{fig:Corner_EZ_Nominal} 
\end{figure*}

The equatorial zone in \Cref{fig:NH3cut-PJmean,fig:NH3-Comb_v1} and  shows a peculiar structure with ammonia increasing with altitude {from $\sim$50 bar to the cloud deck}, in stark contrast with the monotonically decreasing ammonia mixing ratio at all other latitudes. The retrieval favors a solution where the ammonia abundance peaks in the upper troposphere, just below the NH$_4$SH cloud. While a similar structure was found by \cite{Li2020}, we want to show the fidelity of this solution. {We test our assumptions and limitations by expanding both the parameter space and the observational constraints. In our atmospheric fit, we add the zonally averaged VLA observations between 3 and 25 GHz \citep{dePater2016,dePater2019} probing the troposphere down to 10 bars (see \Cref{fig:WF}). For the MWR uncertainties we select the 3-$\sigma$, which compromises between the limb-darkening measurements and the brightness temperature measurements. 
In terms of the parameters, we use the stochastic model, which can better resolve vertical variations in the atmosphere than our thermo-chemical equilibrium model that only fits a single mixing gradient. This is especially relevant for the presence of potential fine scale structure, such as introduced by a cloud precipitation and evaporation feedback cycle below the clouds. Since we introduce a regularization constraint on the vertical variations, we first test the effect of the regularization constraints. In \Cref{fig:EZ_regularization} we show the results of three MCMC runs that have different levels of regularization (Panel a - none, b - mild, and c -moderate). We took 5000 samples from our MCMC run and compute the statistics from them, where the center line corresponds to the mean and the shaded region indicates the 0.16 and 0.84 quantiles. While the regularization changes the small scale structure by smoothing the profile, the overall increase in ammonia is very robust to the regularization constraint. Furthermore the small scale structure, such as the local decrease of ammonia between 10 and 20 bar as seen in the retrievals without constraint is well within the uncertainties of the retrieval and as such not very well constrained by the observations. Therefore, we favor the moderate regularization as it produces a smoother atmosphere when we relax other assumptions. Since the mixing gradient is pretty continuous across the whole domain we average the mixing gradient and show corner plots for this retrieval in \Cref{fig:Corner_EZ_Nominal}, where we show clearly that the ammonia abundance is increasing with decreasing altitude (i.e. a negative mixing gradient)and we exclude solutions with decreasing or constant ammonia abundances with altitude. }

\begin{figure*}[!h]
\centering
\includegraphics[width=0.75\textwidth]{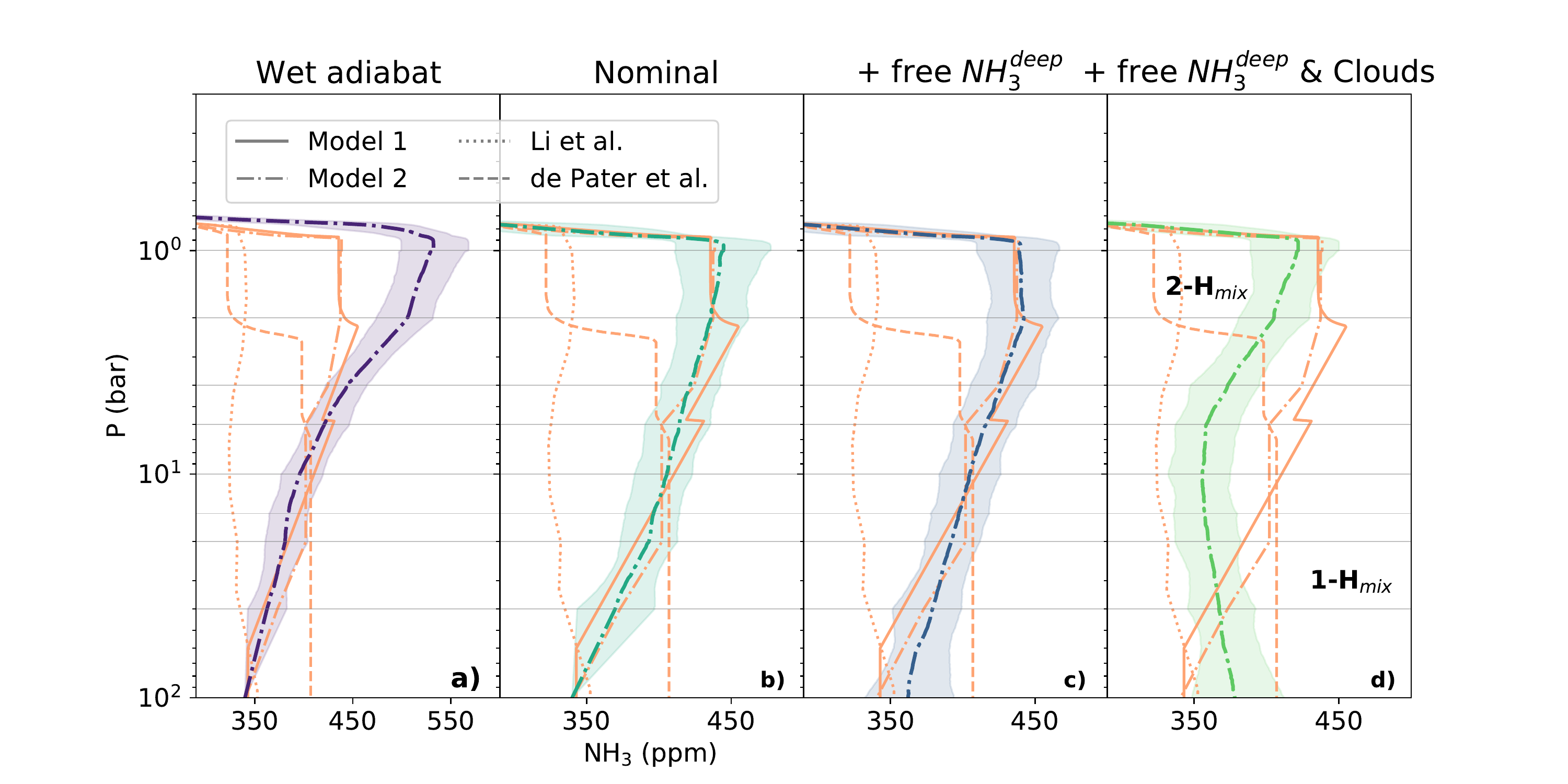}
\caption{{Testing the robustness of the vertical structure against various assumptions in the retrieval compared to various other retrievals (different line styles). Model 1 and Model 2 are slices through \Cref{fig:NH3-Comb_v1}, the ammonia abundance map based on the joint retrieval of the combined VLA and Juno MWR dataset. In panel a we retrieve the atmospheric structure for a wet-adiabatic temperature profile (see \Cref{fig:TPprofile}), and show a large increase of the ammonia abudance (note the different scale for this panel). We compare this to the vertical profile of the nominal dry adiabat in the panel b.  The panel c relaxes the assumption that the atmosphere approaches a globally well-mixed ammonia abundance and instead we retrieve the deep ammonia abundance simultaneously. The right most panel shows the effect of including absorption by cloud particles.}}
\label{fig:NH3cut-EZmodels}
\end{figure*}

\begin{figure}{}
  \centering
  \includegraphics[width=0.5\linewidth]{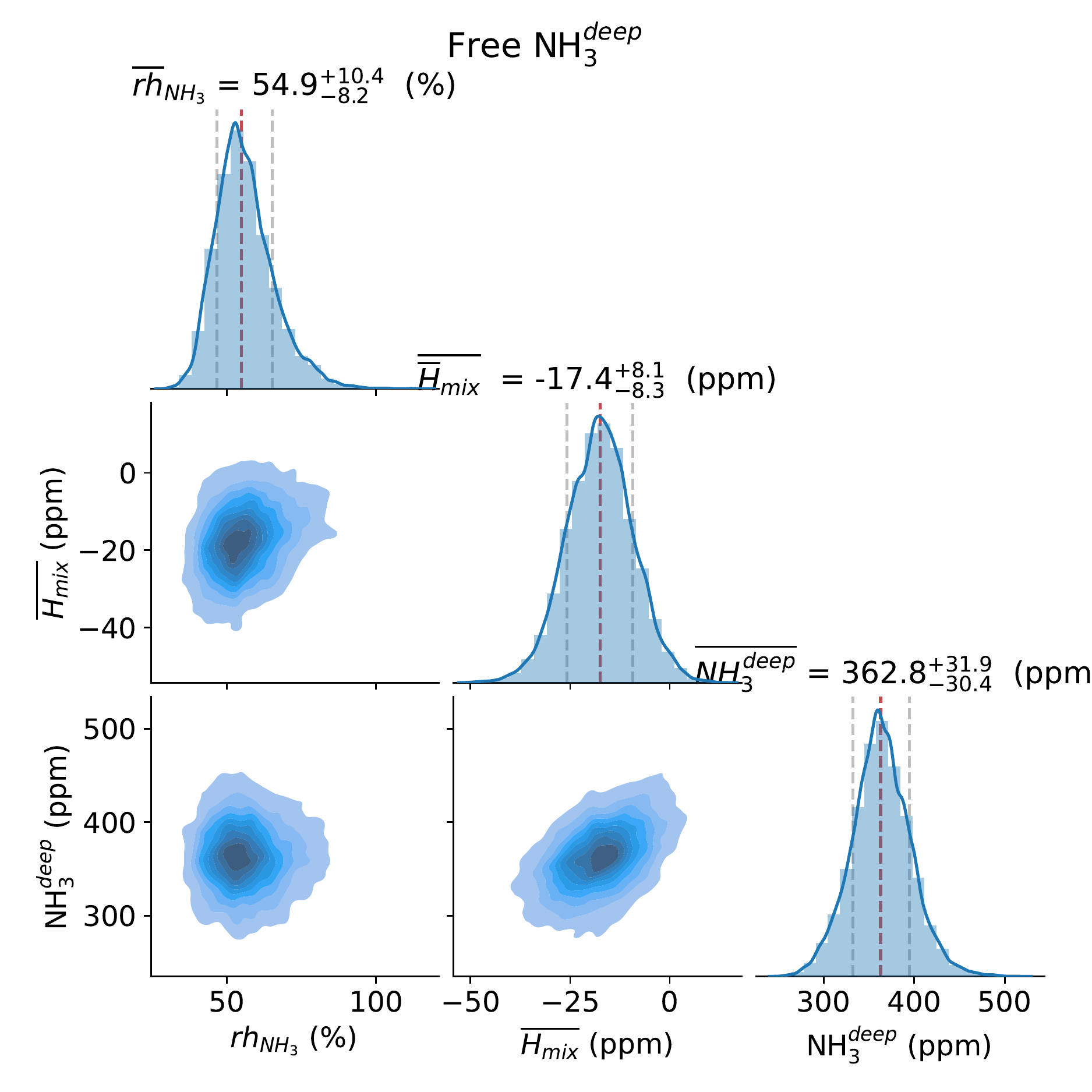}
  \caption{Corner plots for the equatorial inversion corresponding to the profiles shown in \Cref{fig:NH3cut-EZmodels}. In this retrieval we relaxed the assumption that the ammonia is well-mixed and allow the deep ammonia to vary in the fit.}
  \label{fig:Corner_EZ_freeNH3}
\end{figure}

{As explained in Section \ref{ssec:TPprofile}, we assume a temperature pressure profile and attribute variations in the observed quantities to variations in the ammonia abundance. The Voyager radio observations showed that the EZ had a higher temperature than the NEB, so we test the sensitivity of our retrieved ammonia abundance to the temperature structure. The ammonia structure in \Cref{fig:NH3cut-EZmodels}-panel a shows the MCMC results compared to the results based on a dry adiabat in panel b, where the shading indicates the 0.16 and 0.84 quantiles based on the last 5000 samples of our runs. The profile for a wet adiabat differs above $\sim$6 bar, the condensation pressure of water, which decreases the adiabatic lapse rate. This higher temperature of the wet adiabat is compensated by higher ammonia abundances above 6 bars to match the MWR and VLA observations.
Next we relax the assumption that the atmosphere approaches a state of being well-mixed globally deep in the atmosphere and we allow the deep ammonia abundance to vary along with other parameters in the model, instead of using the deep ammonia abundance as an a-priori constraint. Panel c of \Cref{fig:NH3cut-EZmodels} shows the results of MCMC simulations. We plot the distribution of the final parameters in \Cref{fig:Corner_EZ_freeNH3} to show how well constrained the parameters are. While the top of the atmosphere remains unchanged, the mixing gradient is slightly decreased which is compensated by an increase in the deep ammonia abundance by 22 ppm compared to the global mean of 340 ppm, which is still within the errorbars of our retrievals (see Section \ref{ssec:DA}).}

\begin{figure}{}
  \centering
  \includegraphics[width=0.5\linewidth]{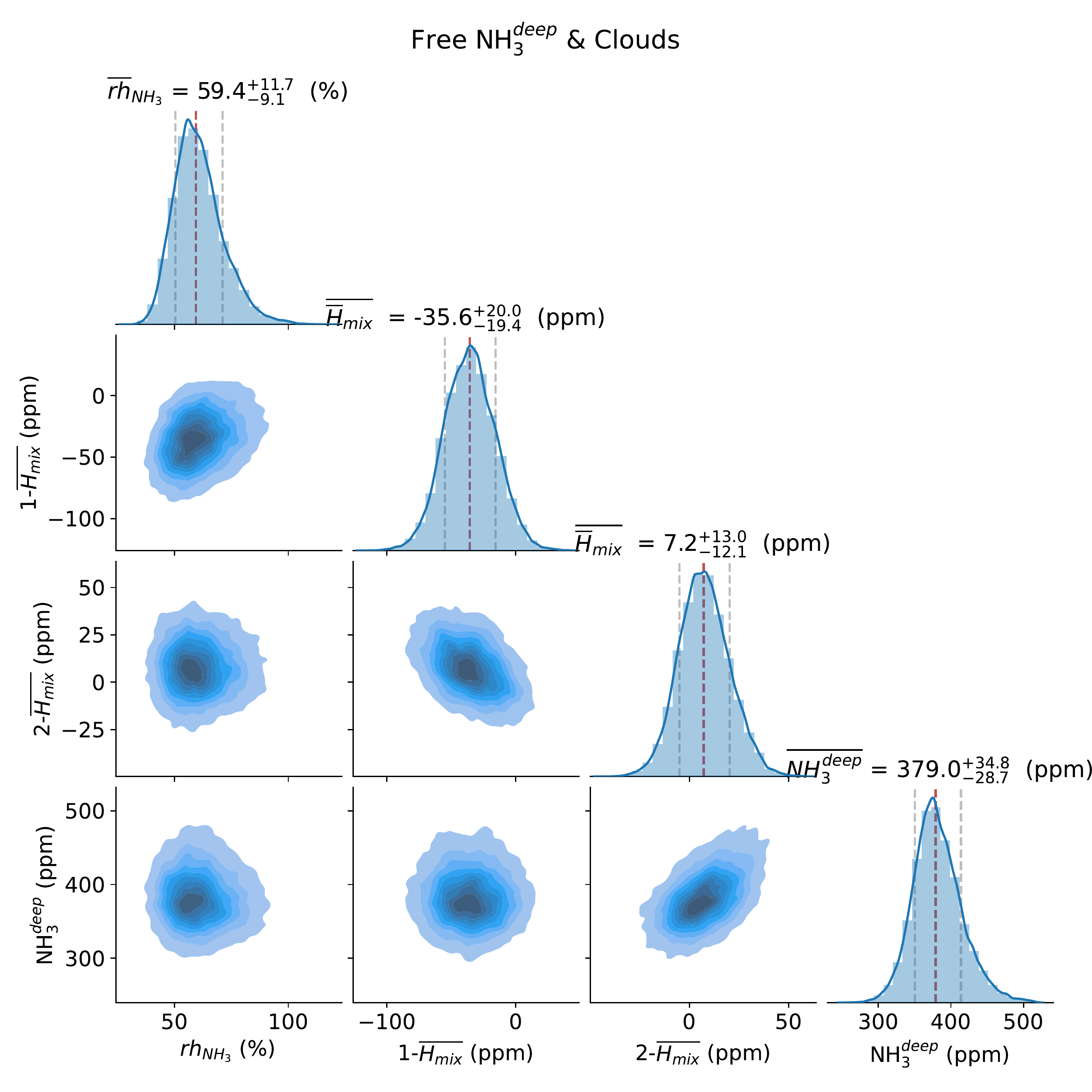}
  \caption{Corner plots for the equatorial inversion including the effect of radio absorption by cloud particles. The mixing gradients correspond to 1-H to mixing above 6 bar, and 2-H corresponds to mixing below. }
  \label{fig:Corner_EZ_freeNH3+clouds}
\end{figure}%


One of the big assumptions is that effect by clouds can be ignored. To test the validity of this assumption we include absorption by clouds based on the parameterization by \citet{dePater1985,dePater2005}. Cloud densities in these models are computed like adiabatic liquid water contents for terrestrial clouds, i.e., they represent maximum cloud densities based upon assuming no loss by precipitation. The results for the atmosphere are shown in panel d of \Cref{fig:NH3cut-EZmodels}, where he ammonia abundance profile has clearly changed significantly compared to the previous models. Below the water condensation level the atmospheric structure is consistent with a constant abundance (1-H mixing gradient), with a change in gradient above the water cloud level signified by a sharp increase in the ammonia abundance (2-H mixing gradient) consistent with a rain-evaporation cycle. The retrieved deep ammonia abundance (see \Cref{fig:Corner_EZ_freeNH3+clouds}) of 379 ppm is just outside of the $1\sigma$ limit of the global ammonia abundance that we retrieved without considering the opacity by clouds. This structure is most consistent with physical models, however beckons the question what other impact the poorly known cloud particles may have on the retrieved atmospheric parameters.

\subsection{Mid-latitude structure}\label{ssec:ML}
\begin{figure*}[!h]
\centering
\includegraphics[width=0.75\textwidth]{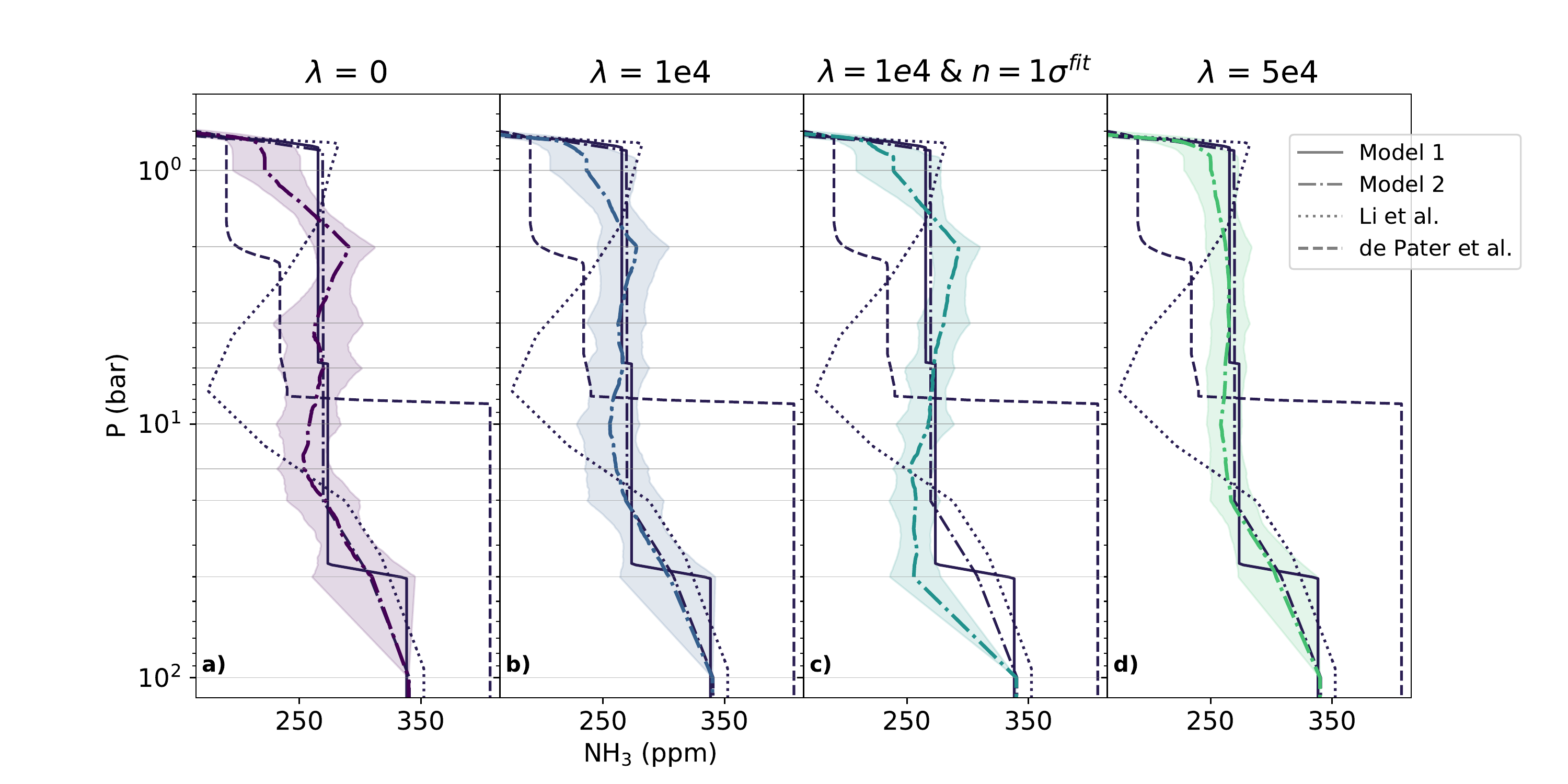}
\caption{Testing for the inversion signal at midlatitudes as seen by \citet{Li2017}, by retrieving an atmosphere based on both VLA and Juno data. We test the effect of vertical regularization on the mid latitude structure to show the confidence in the retrieved mid-latitude structure. The panels correspond to no (panel a), mild (panel b) and moderate vertical regularization (panel d). The various line style corresponds to the various retrievals, with the dot-dashed line and the shading indicated the 0.16 and 0.84 quantiles of the MCMC results. Panel c uses mild vertical regularization but reduces the uncertainty fidelity to n = 1. Regardless of chosen regularization we cannot reproduce the magnitude of the inversion as seen by the dotted line.}
\label{fig:ML_regularization}
\end{figure*}

{The presence of the inversion at pressures around ~6 bar (see \citet{Li2017})  has lead to much speculation on the physical processes that could cause such a structure. A possible explanation would be a set of ferrel cells extending polewards that alternately enrich and deplete the  atmosphere \citep{Duer2021}. While our thermo-chemical equilibrium model with a singular mixing gradient would not be able to resolve such a structure, our stochastic model built upon the framework by \citet{Li2017} should be able to find such a structure. In our optimizer results we could not find any consistent evidence for the depletion as seen by \citet{Li2017} (see \Cref{fig:NH3-PJmean,fig:NH3-Comb_v1}). The biggest factor in controlling this structure is the amount of regularization we introduce for the vertical gradients, so that we tested our retrieval against various regularization constants $\lambda$. Additionally, we increased the constraints by adding the VLA information to use all available observations.  We chose -30 degree latitude where \citet{Li2017} found a strong inversion signal with the ammonia dropping down to 175ppm compared to 275ppm at the ammonia ice cloud level. In \Cref{fig:ML_regularization} we show the results for various vertical regularization constants on the retrieved structure, similarly to \Cref{fig:EZ_regularization}.  Panel a shows the resulting atmospheric structure for a retrieval with no vertical regularization, while panel b and d correspond to mild and moderate. Regardless of regularization constant we cannot reproduce the inversion at $\sim$ 6 bar (dotted line in \Cref{fig:ML_regularization}.) We also test if our choice of uncertainty (n=3$\sigma^{fit}$) could influence the results. In panel c we show the results of using n=1 for the fit uncertainty, i.e. an increased influence of the limb-darkening parameter on the retrieval. The resulting structure develops a slight negative gradient, however, no consistent inversion. This agrees with the conclusion by \citet{dePater2019}, who showed that ammonia abundance profiles by \citet{Li2017} gave an inferior fit to the data as compared to their derived abundance profile. Similarly, \citet{Ingersoll2021} used Perijove 1-7 observations and did not find a structure consistent with an inversion at mid latitudes. }

\section{Conclusions}
Radio observations allow to disentangle some of the mysteries that Jupiter still holds for us. A combined analysis of VLA and Juno observations allow us to constrain the vertical structure of the atmosphere; however, this is not unique due to the fact that radio waves originate from a range of pressures. We introduce disequilibrium factors into a thermo-chemical equilibrium model \citep{Weidenschilling1973,Atreya1985} that describe the distribution of trace gases and use these factors in our inversion technique. Fitting the required quantities, however, also requires a good understanding of the fidelity of the data product. For this reason, we developed a data reduction pipeline, that obtains the brightness temperature and limb-darkening coefficient along with their uncertainties. Coupling these two, results in a distribution of trace gases (see \Cref{fig:NH3-PJmean}), the sensitivity of our solutions to the fit uncertainties (see \Cref{fig:PJmean_params}) and our certainty in the derived atmospheric distribution (see \Cref{fig:corner_EZ,fig:corner_NEB}). \\ 

Our approach recovers an atmosphere that confirms the structure that was seen by the VLA in the past \citep{dePater1986,dePater2019} and by Juno \citep{Li2017}, however, we also find some remarkable differences. We find that the deep ammonia abundance of $340.5^{+34.8}_{-21.2}$ ppm  ($2.30^{+0.24}_{-0.14} \times$ solar abundance) is lower than the Galileo \citep{Wong2004} and VLA results \citep[which were based on the Galileo results]{dePater2016,dePater2019}, and confirm  the Juno results \citep{Li2017}. The atmosphere only approaches a state of being well-mixed at pressures greater than {~30} bars, sitting under a layer between 20 and 30 bar in which the ammonia abundance decreases with increasing altitude. The atmospheric structure requires processes that deplete the atmosphere down to such high pressures, or alternatively require a mechanisms that locally enrich the atmosphere, such as in the equatorial zone. The systematic deeper depletion of ammonia abundance in the zones compared to the belts supports the existence of a vertically stacked circulation structure that is constrained to the top ~20-30 bars of the atmosphere. At the top of troposphere the formation of the NH$_4$SH and ammonia cloud layers dominates the vertical structure in the ammonia abundance.

The uncertainties that we retrieve from the raw Juno observations allow us to study the sensitivity of our results to these observational uncertainties. Especially the low frequency observations are hard to fit, where the limb-darkening coefficient and the brightness temperature favor different solutions for the structure of the atmosphere at altidues below 10 bar. Higher up in the atmosphere the uncertainties have a very small effect on the retrieved atmosphere (see \Cref{fig:DA}). \\ 
The equatorial zone represents an interesting diversion from the general atmosphere. It shows the highest ammonia abundance on the planet, with an indication that the ammonia is increasing with altitude, in stark contrast to the rest of the planet. This signal is robust to the chosen uncertainties (see \Cref{fig:PJmean_params}) and can be reproduced even when we loosen the assumption and let the deep ammonia vary. This structure can also be reproduced when adding the VLA observations (see \Cref{fig:NH3cut-EZmodels}) and indicates an increase in ammonia over a large pressure range. Interestingly, when adding absorption by clouds the structure is consistent with an increase starting only at the pressure of the water clouds.\\ 
So how can we advance our understanding of the deep atmosphere? First and foremost, we need to investigate the signal in Channel 1 which so far has been mostly ignored. With the differences between the various ammonia absorption coefficients at high pressures \citep{Hanley2009,Bellotti2016,Bellotti2017} alone, Jupiter's deep atmosphere remains poorly understood. Furthermore, at low frequencies the absorption by water clouds \citep{dePater2005} and other species such as alkalis \citep{Bellotti2018} becomes significant, both of which are poorly understood and often ignored in the RT models. Both lab measurements at high pressures and temperature and ab-initio calculations are essential to fully model these contributions to the opacity. The other big unknown is the temperature structure below the clouds. While mid-infrared observations can probe down to $\sim$0.8 bar, the structure below is largely unconstrained, and usually assumed to be adiabatic (wet or dry). Remote observations cannot retrieve the temperature structure directly; understanding atmospheric dynamics better can help construct more realistic atmospheric temperature profiles. 

\acknowledgments
We would like to thank the Juno team for encouraging an independent analysis of the MWR data and for making their raw data available on the PDS. We would like to thank Ned Molter and Josh Tollefson for their thoughtful comments on the difficulty of interpreting radio data. We thank Bryan Butler and Bob Sault for their help on producing the map in Figure 1. And lastly, a big thank you to the two referees and the lively discussions about this work. 
The VLA data used in this report, associated with project code 13B-064 (2013-2014) and 16B-048 (2016), are available from the NRAO Science Data Archive at \url{https://archive.nrao.edu/archive/advquery.jsp}. The National Radio Astronomy Observatory is a facility of the National Science Foundation operated under cooperative agreement by Associated Universities, Inc. This research has been supported by NASA’s Solar System Observations (SSO) award 80NSSC18K1001 to the University of California, Berkeley. 

\section*{Appendix A - Beam-Convolved Emission Angle }\label{sec:ApA} 
\begin{figure}{}
  \centering
  \includegraphics[width=0.5\linewidth]{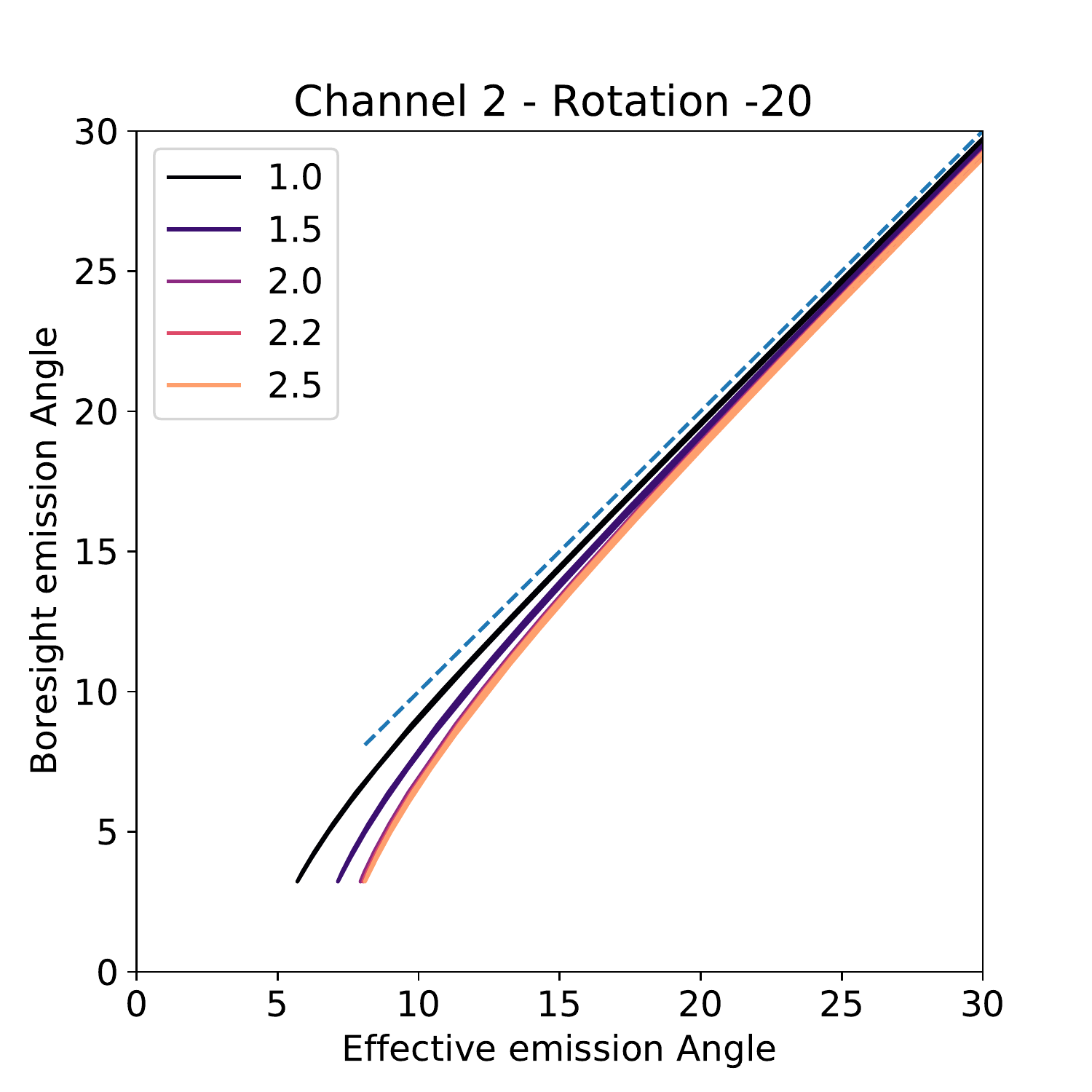}
  \caption{Relationship between the beam-convolved emission angle and boresight emission angle. The different colors indicate the integration region in units of the HPBW. Further away from the beam center, the beam sensitivity is dropping rapidly, and the contribution to the averaging becomes small. Therefore, the results converge for sufficiently large integration regions. Our final results are based on 2.5 HPBW, which contains 99\% of the emission. This example is for Perijove 1 - Channel 2 observations just before closest approach to the planet. }
  \label{fig:eea_rot-20C2}
\end{figure}

\begin{figure}{}
  \centering
  \includegraphics[width=0.5\linewidth]{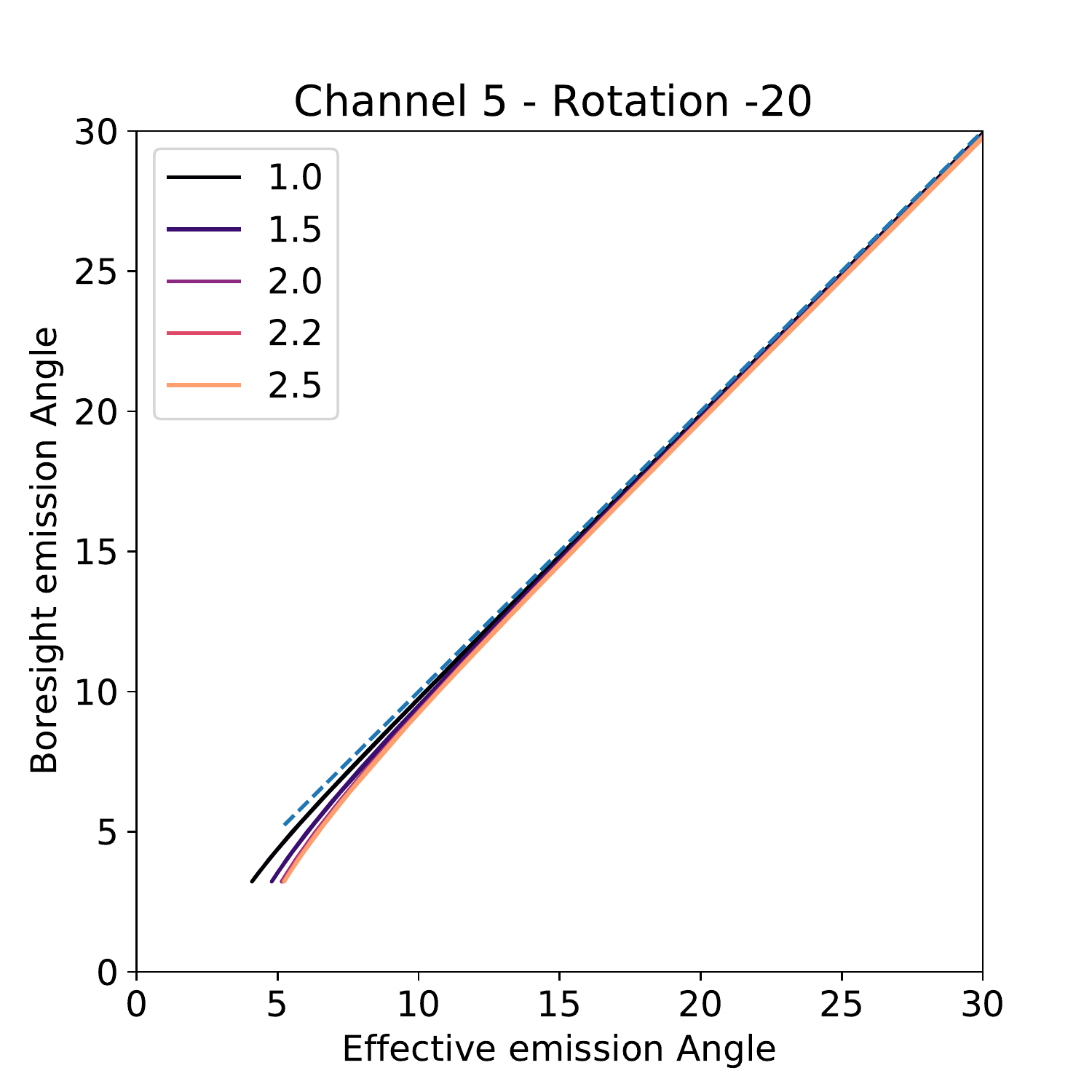}
  \caption{Same as \Cref{fig:eea_rot-20C2} but for Channel 5.}
  \label{fig:eea_rot-20C5}
\end{figure}

As explained in Section \ref{sec:data}, for each observation we compute the beam-convolved emission angle to account for the large range of emission angles within the beam. Since we are interested in the relationship between the emission angle and the brightness temperature we must assure that the emission angle is representative of the observations. Especially at low emission angles, the extended beam covers a wide range of emission angles, and the boresight emission angle is not an adequate indicator for the observation. \Cref{fig:eea_rot-20C2,fig:eea_rot-20C5} shows the relationship between the two quantities for a full rotation of the spacecraft. The large beam-size of Channel 2 makes the effect more pronounced than the observations in Channel 5.

\section*{Appendix B - Synchrotron filter }
\begin{figure*}[h]
\centering
\includegraphics[width=0.8\textwidth]{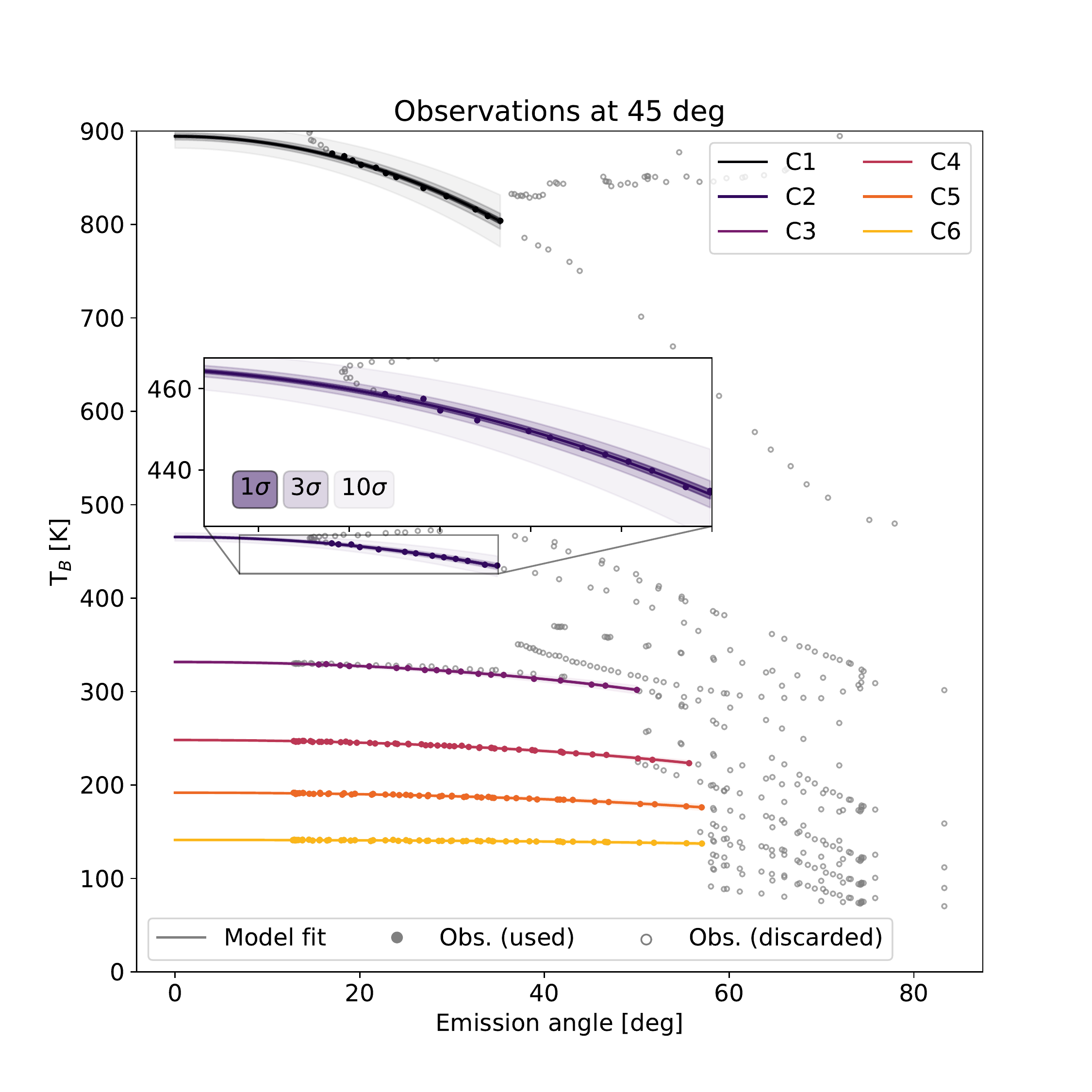}
\caption{ Same as \Cref{fig:TBandLDfit} but for a latitude where the synchrotron contribution is noticeable. Note that the there are two branches in Channel 1 and 2 that have a large difference in the temperature for the same emission angle. The increase temperature is due to the synchrotron leaking in through the side lobes. Our spatial restriction can handle the synchrotron contribution well, ignoring the upper, contaminated branch of the observations.   }
\label{fig:PJ3_syncfilt}
\end{figure*}

To best understand the spatial synchrotron filter, in \Cref{fig:PJ3_syncfilt} we show a region of the planet where the radiation from the main synchrotron belt is leaking in when looking towards the equator. Compared to \Cref{fig:TBandLDfit}, where the measurements all fall onto the same curve, we can see that in regions with heavy syynchrotron contribution, there are two branches to the radiation. We have separated out the signal that is looking towards the equator (upper branch of empty circles), while the useful data are the ones looking towards the poles of the planet. Even at low emission angle the impact of the synchrotron radiation can be large.

\bibliography{main}{}
\bibliographystyle{aasjournal}



\end{document}